\documentclass[final,3p,times]{elsarticle}
\usepackage{graphicx}
\usepackage{wrapfig}
\usepackage{amssymb}
\usepackage{mathrsfs}
\usepackage{amsmath}
\usepackage{comment}
\usepackage{amsbsy}
\usepackage{mathrsfs}
\usepackage{amsfonts}
\usepackage{epsf,color,colordvi,pifont}
\usepackage{lineno}
\usepackage{nicefrac}
\usepackage[notref,notcite]{showkeys} 
\def\deltabar{{\mathchar'26\mkern-9mu\delta}}
\def\dbar{{\mathchar'26\mkern-12mu d}}
\DeclareFontFamily{OT1}{pzc}{}
\DeclareFontShape{OT1}{pzc}{m}{it}{<-> s * [1.10] pzcmi7t}{}
\DeclareMathAlphabet{\mathpzc}{OT1}{pzc}{m}{it}

\usepackage{amsmath}

\def\dbar{{\mathchar'26\mkern-12mu d}}
\journal{Nuclear Physics B}
\begin{document}
\begin{frontmatter}
\title{Laser-driven search of axion-like particles  including   vacuum polarization  effects}

\author{S.  Villalba-Ch\'avez}
\ead{selym@tp1.uni-duesseldorf.de}
\address{Institut f\"{u}r Theoretische Physik I, Heinrich Heine Universit\"{a}t D\"{u}sseldorf\\ Universit\"{a}tsstr. 1, 40225 D\"{u}sseldorf, Germany}
\begin{abstract}
Oscillations of photons into axion-like particles  in  a high-intensity  laser field  are  investigated. Nonlinear QED effects
are considered  through the low energy behavior of the vacuum polarization tensor, which  is derived  from the  Euler-Heisenberg Lagrangian
in the one-loop and weak field approximations.  The expressions  obtained in this framework  are  applied to  the configuration
in which the strong background  field is a circularly  polarized monochromatic plane wave. The outcomes of this  analysis  reveal  that, in
the regime of  low energy-momentum transfer, the axion field  induces  a  chiral-like birefringence and  dichroism in the vacuum which is not manifest  in a pure QED  context.
The corresponding  ellipticity and  angular  rotation  of the polarization plane are also determined. We take advantage of  such observables  to  impose exclusion limits on the
axion  parameters.  Our  predictions  cover axion masses for  which a setup  based on  dipole magnets provides less stringent constraints. Possible experimental scenarios in which  our
results could be tested  are also discussed.
\end{abstract}
\end{frontmatter}

\section{Introduction}

The  nonlinear vacuum of Quantum Electrodynamics (QED) is  an illuminating laboratory for exploring  physics  beyond the framework of the
Standard  Model (SM) of  fundamental interactions. Over the last few years  there  have been substantial efforts devoted to  employ
its unconventional properties  in the search of  a  plausible but   elusive  pseudo-scalar particle known as the axion.
This  hypothetical Nambu-Goldstone  boson emerges  from the spontaneous breaking of the Peccei-Quinn symmetry and  turns out to be a
distinctive quantity  within the  solution to the strong CP problem \cite{Peccei:1977hh,Wilczek:1977pj,Weinberg:1977ma}. It additionally  conforms to   the
paradigm  of   axion-like particles (ALPs) closely  associated with some extensions of the SM which naturally emerge from  string compactifications \cite{Witten:1984dg,Lebedev:2009ag}.
Conceptually the ALPs encompass both scalar and pseudo-scalar bosons \cite{Gies:2007ua,Biggio:2006im,Gabrielli:2006im} being likely    candidates
for the dark matter  of the universe \cite{Duffy:2009ig,Sikivie:2009fv,Raffelt:2006rj,Baer:2010wm}. Their  conversion   into an electromagnetic
field is  a long-standing prediction   which  has  been frequently analyzed in a constant magnetic field  \cite{Sikivie:1983ip,Sikivie:1985yu,maiani,Raffelt:1987im}. 
In  this external field configuration  the absorption of a photon into a real ALP  induces an attenuation of a probe laser beam.
Since the  amount of absorbed  photons is different for each propagating mode the vacuum behaves as a dichroic medium.  Simultaneously, in the presence of a external magnetic field, the ALP-photon coupling
modifies the vacuum birefringence   caused    by  the   polarization of  virtual electron-positron pairs \cite{Dittrich,adler,shabad2,shabad4}. 
Both phenomena have inspired  polarimetric experiments   in which  indirect evidence of  ALPs   could be detected.  Among the most  significant
collaborations are  BFRT \cite{Cameron:1993mr},  PVLAS \cite{Zavattini:2007ee}, BMV \cite{BMVreport} and Q$\&$A \cite{Chen:2006cd}.
On the other hand, there exists  another interesting mechanism  of  finding traces of the ALPs existence which relies on  the photon regenerative property,
commonly known as  ``Light Shining Through a Wall''   \cite{VanBibber:1987rq,Adler:2008gk,Arias:2010bh,Redondo:2010dp}.
This  has  been experimentally  implemented in several  collaborations  such as ALPS \cite{Ehret:2010mh,Ehret:2009sq}, GammeV \cite{Chou:2007zzc,Steffen:2009sc},
LIPSS  \cite{Afanasev:2008jt}, OSQAR \cite{Pugnat:2007nu} and BMV \cite{Robilliard:2007bq,Fouche:2008jk}. However,
despite the push to  detect  these particles,  the  results provided by  both  kinds of experiments are far from  proving  that the  photon
oscillations into ALPs occur.   Instead, upper bounds on the unknown parameter of   ALPs, i.e.,  coupling  constant $g$ and  mass $m$  have been established, as well as  for  other weakly interacting
particles including  paraphotons \cite{Okun:1982xi,Masso:2006gc,Ahlers:2007rd,Ahlers:2007qf,Goodsell:2009xc}  and  mini-charged  particles \cite{Dudas:2012pb,Holdom:1985ag,Gies:2006ca,Jaeckel:2009dh,
Dobrich:2012sw,Dobrich:2012jd}.   The main difficulty  in these experiments stems from the  projected  lightness of the
ALPs  and   the  weakness of their  coupling constants, hence the detection of  their  tiny
observable effects represents a huge technical challenge.

An optimal  setup is necessary to overcome this obstacle.  Very often  the  magnetic field strength $\vert\pmb{B}\vert$  as well as its spatial extension  $\ell$ is  exploited  to partially achieve
this goal. Their combined effects, usually evaluated  through the product $\vert\pmb{B}\vert \ell$,  facilitate  the enhancement of observables associated with the mixing process as long as both
quantities are increased.  Frequently, in high-precision optical experiments,   field strengths of the order of  $\vert\pmb{B}\vert\sim\mathpzc{o}(10^4-10^5)\ \rm G$ are   extended  over lengths
$\ell\sim\mathpzc{o}(10^2-10^3)\ \rm  cm$ so that $\vert \pmb{B}\vert\ell\sim \mathpzc{o}(10^6-10^{8})\ \rm G cm$.  Although the incorporation of interferometric techniques  has allowed to extend the  interaction region up to   macroscopically
distances $\ell\sim\mathpzc{o}(10^3)\ m$,  the  attainable laboratory  values of $\vert\pmb{B}\vert$  are not strong enough to  manifest the desirable effects.
Gradually, the technology of high-intensity lasers   is proving  to be an alternative tool as it can  achieve   much stronger  field strengths
$\vert\pmb{B}\vert\sim\mathpzc{o}\left(10^9\right)\  \rm G$ in a short  space-extension of the orders of $\ell\sim \mathpzc{o}(1-10)\ \mu\rm m$ allowing for the product
$\vert\pmb{B}\vert \ell\sim \mathpzc{o}(10^{5}-10^6)\ \rm Gcm$. However, this tiny  interaction region  could be  compensated for   by the envisaged ultrahigh intensities at future 
laser facilities.  Contemporary projects such as  the Extreme Light Infrastructure (ELI)
\cite{ELI} and the Exawatt Center for Extreme Light Studies (XCELS)  \cite{xcels}  are being   designed to reach the unprecedented  level  of
$\vert\pmb{B}\vert\sim\mathpzc{o}(10^{12})$ G, an order  of magnitude   below  the critical magnetic  field of QED $B_c=4.42\times10^{13}$ G,
above  which  the superposition principle is no longer valid and  the product $\vert\pmb{B}\vert\ell\sim \mathpzc{o}(10^{8}-10^9)\ \rm Gcm$   exceeds by an order of magnitude the maximum value resulting
from experiments driven by a constant magnetic field.  This  has raised  hopes that nonlinear effects  including  vacuum
birefringence \cite{Heinzl1,Heinzl}, photon splitting  \cite{DiPiazza:2007yx},  diffraction effects \cite{DiPiazza:2006pr,BenKing:2009,BenKing:2010,Tommasini:2010fb,Hatsagortsyan:2011bp,Gies:2013yxa}
and the spontaneous production of electron-positron pairs  from the vacuum  \cite{Hebenstreit:2009km,ruf2009zz,mocken2010}  may soon be within an
experimental scope  with  purely laser-based setups. There has been  some important  progress  within the field of  ALPs: some estimations have been put forward in
\cite{mendonza,Gies:2008wv} and,  recently, a more  in-depth   investigation has established stringent constraints on the  coupling constant
in  regions of axion masses for  which a laboratory setup   based on  dipole magnets provides less severe limits \cite{Dobrich:2010hi,Dobrich:2010ie}.

Due to   these considerations  and motivated  by the  theoretical relevance  of the ALPs,  it is of  interest to improve  our  understanding
of  photon-ALP(s) and ALP(s)-photon conversion  in an experimentally attainable setup in which a  high-intensity  laser wave   is  taken as the background
external field of the theory.  This work contributes to this endeavor by focusing on the phenomenological aspects associated with  pseudoscalar ALPs in the 
field of a circularly polarized monochromatic plane wave.  Our main purpose is to  explore  the effects of these pseudoscalar  particles on  physical
observables which can be used to improve the exclusion limits on its mass and coupling constant.  To this end,  we have organized the  paper in the following
form: in Sec.~\ref{svpfgpw}  the equations of motion  associated with the oscillations processes are derived  in the field of a  plane wave of arbitrary shape.
In addition, the low energy behavior of the  vacuum polarization tensor is obtained from the  Euler-Heisenberg Lagrangian in the one-loop and weak
field approximations.  This is followed  by a particularization of the problem to the case in which the strong laser field is  circularly polarized monochromatic plane wave.
In Sec.~\ref{circulaexternafield}, the corresponding dispersion relations and  the equations of motion of fields involved in the Lagrangian   are solved.  This 
setup reveals that--contrary to what occurs
in a pure QED context--chiral  birefringence and dichroism of the vacuum  are  induced  by the ALP-photon coupling.
In Sec.~\ref{section4}  the  observables associated  with   polarimetry techniques  are  derived  and exclusion limits are then established. Finally,  we  present a
summary and outlook of our research work.

\section{Photon-Axion mixing  in  the field of a plane wave of arbitrary shape \label{svpfgpw}}

Nonlinear effects of the electromagnetic field emerge as a consequence of  effective  couplings   provided by the polarization
 of  virtual electron-positron pairs. For small   energy-momentum transfer,  below  the energy scale specified  by the electron
mass $m_0$,  the physical phenomena associated with this theory can be described in a unitary way by means of  the
Euler-Heisenberg Lagrangian \cite{euler,Schwinger}. For field strengths  much  weaker than the corresponding  critical electric and magnetic
fields, the leading  behavior of this Lagrangian turns out to be\footnote{From now on  natural and Gaussian
units  $4\pi\epsilon_0=\hbar=c=1$ will be used.}
\begin{eqnarray}\label{EHL}
\mathfrak{L}=-\frac{1}{4\pi}\mathfrak{F}+\frac{1}{8\pi}\mathfrak{L}_{\mathfrak{F}\mathfrak{F}}\mathfrak{F}^2+\frac{1}{8\pi}\mathfrak{L}_{\mathfrak{G}\mathfrak{G}}\mathfrak{G}^2,
\end{eqnarray} where the quadratic terms in the field invariants  $\mathfrak{F}=\frac{1}{4}F_{\mu\nu}F^{\mu\nu}$ and
$\mathfrak{G}=\frac{1}{4}\tilde{F}_{\mu\nu}F^{\mu\nu}$ account for the  quantum corrections to the Maxwell Lagrangian
$\mathfrak{L}_{M}=-\frac{1}{4\pi}\mathfrak{F}$, with $F_{\mu\nu}$ the electromagnetic field tensor and
$\tilde{F}_{\mu\nu}=\frac{1}{2}\varepsilon_{\mu\nu\sigma\beta}F^{\sigma\beta}$ its dual.  In the one-loop approximation,
their respective coefficients are given  by
\begin{equation}\label{coefficients}
\mathfrak{L}_{\mathfrak{F}\mathfrak{F}}=\frac{4}{45}\frac{\alpha}{\pi}\frac{e^2}{m_0^4} \qquad \mathrm{and} \qquad \mathfrak{L}_{\mathfrak{G}\mathfrak{G}}=\frac{7}{45}\frac{\alpha}{\pi}\frac{e^2}{m_0^4},
\end{equation} with $\alpha=e^2\approx1/137$  the fine structure constant 
and   $e$  the  absolute value of the electron charge.

The incorporation of an interacting pseudoscalar sector is usually done by preserving the fundamental symmetries of QED.
In line with this assumption,  the nonlinear effective action which describes the minimal coupling between the photon  field $A_\mu(x)$ and an ALP $\phi$  reads
\begin{equation}\label{actionpreliminar}
\mathcal{S}=\int d^4 x\left\{\mathfrak{L}+\frac{1}{2}(\partial_\mu\phi)^2-\frac{1}{2}m^2\phi^2+\frac{g}{4\pi}\phi\mathfrak{G}\right\},
\end{equation} where $m$ and $g\sim1/\Lambda$ are the mass and  coupling constant of the ALP, respectively. Here  $\Lambda$ is a  parameter with
 dimension of energy, which for a  QCD axion  represents  the phenomenological energy scale at which the Peccei-Quinn symmetry  is broken
\cite{Peccei:1977hh}.

\subsection{Low energy behavior of the vacuum polarization tensor}

Since we are interested in  analyzing  how this coupling modifies the propagation  of  small-amplitude electromagnetic waves $a_\mu(x)$ in  an
external  background field $\mathscr{A}_\mu(x)$, it is convenient to express $A_\mu(x)=\mathscr{A}_\mu(x)+a_\mu(x)$ and  expand  $\mathcal{S}$
in power series of  $a_{\mu}(x)$ above   $\mathscr{A}_\mu(x)$. This procedure leads to the functional action
\begin{eqnarray}\label{EASCPP}
\mathcal{S}[a,\phi]=\int d^4x\left\{\mathscr{L}-\frac{1}{2}\phi\left(\square+m^2\right)\phi+\frac{g}{8\pi}\phi\tilde{\mathscr{F}}_{\mu\nu}f^{\mu\nu}\right\},
\end{eqnarray}where $f^{\mu\nu}=\partial^\mu a^\nu-\partial^\nu a^\mu$ and $\mathscr{F}^{\mu\nu}=\partial^\mu \mathscr{A}^\nu-\partial^\nu \mathscr{A}^\mu$
are the electromagnetic tensors for  small-amplitude waves  and  strong laser field, respectively. Here $\square\equiv\partial^2/\partial t^2-\nabla^2$,  and
 \begin{eqnarray}\label{Lagragiansdddad}
\mathscr{L}=\frac{1}{2}\int d^4x^\prime a^{\mu}(x)\mathscr{D}_{\mu\nu}^{-1}(x,x^\prime)a^\nu(x^\prime)
\end{eqnarray} is the quadratic part of the effective Lagrangian in $a_\mu(x)$,  with  $\mathscr{D}_{\mu\nu}^{-1}(x,x^\prime)$
denoting  the inverse  photon  propagator in an external background field. Its general structure can be seen  from the QED Schwinger-Dyson equations
\cite{Dyson:1949ha,Schwinger:1951ex1,Schwinger:1951ex2,Alkofer:2000wg,fradkin} and turns out to be
\begin{eqnarray}\label{propagatorinversoQED}
\mathscr{D}_{\mu\nu}^{-1}(x,x^\prime)=\frac{1}{4\pi}\left[\square \mathpzc{g}_{\mu\nu}-\partial_{\mu}\partial_\nu\right]\delta^{(4)}(x-x^\prime)+\frac{1}{4\pi}\Pi_{\mu\nu}(x,x^\prime).
\end{eqnarray} Here  $\mathpzc{g}_{\mu\nu}$ is the metric tensor whose diagonal components are
$\mathpzc{g}^{11}=\mathpzc{g}^{22}=\mathpzc{g}^{33}=-\mathpzc{g}^{00}=-1.$  Obviously, the first term  in Eq.~(\ref{propagatorinversoQED}) gives  the  Maxwell
Lagrangian while the second   is  responsible for the quantum  corrections which, for small-amplitude electromagnetic  waves, are
described by the vacuum polarization  tensor $\Pi_{\mu\nu}(x,x^\prime).$

To reveal the  low energy behavior of this tensor  and  obtain a clear picture of the photon spectrum it is sufficient  to  variate the action
[Eq.~(\ref{actionpreliminar})]  with respect to $A_\nu(x)$ twice, set the field invariants $\mathfrak{F},$ $\mathfrak{G}$  and   $\phi$ to zero,
and  compare the resulting expression to  Eq.~(\ref{propagatorinversoQED}). The former  evaluation  is in correspondence with the fact that  for
plane waves--crossed field, equal strengths--the field invariants  $\mathfrak{F}$ and $\mathfrak{G}$  vanish identically. In contrast, by setting
$\phi=0$ we are assuming that there is no expectation for the axion fields permeating  the universe, or that such vacuum expectation value  is
neglectable  in comparison with the fluctuations  in  which we are interested\footnote{A nonvanishing expectation value of $\phi$  might play a relevant
role when  axions are considered as dark matter candidates \cite{Duffy:2009ig}.}.  As long as this is the case,   we find  that
\begin{eqnarray}\label{IRVPT}
\Pi_{\mu\nu}(x,x^\prime)=-\left\{\mathfrak{L}_{\mathfrak{F}\mathfrak{F}}\mathscr{F}_{\beta\nu}\mathscr{F}_{\alpha\mu}\partial^\alpha\partial^\beta+\mathfrak{L}_{\mathfrak{G}\mathfrak{G}}\tilde{\mathscr{F}}_{\beta\nu}\tilde{\mathscr{F}}_{\alpha\mu}\partial^\alpha\partial^\beta\right\}\delta^{(4)}(x-x^\prime),
\end{eqnarray}where the following relation $\mathscr{F}^{\mu\nu}\partial_\mu \mathscr{F}_{\lambda\rho}\propto \varkappa_\mu \mathscr{F}^{\mu\nu}=0$ has been used. The procedure used to obtain  the expression above has also  been
successfully applied to the case of constant background fields \cite{Shabad:2011hf,VillalbaChavez:2012ea}. This  is applicable as long as the electromagnetic
field  is slowly varying on a linear spacetime scale of the order of the Compton-wavelength $\lambda_c=1/m_0=3.9\times10^{-11}$cm, otherwise
the spatial  and temporal  dispersion become important issues  and one is forced to consider  the general expression of $\Pi_{\mu\nu}$
calculated from  Feynman diagram  techniques [see  Fig. (\ref{fig:mb000})] in the Furry picture. This calculation was originally carried
out  by Batalin and Shabad  \cite{batalin}  in  the special case  of a constant electromagnetic field.   In contrast, Ba\u{\i}er, Mil'shte\u{\i}n and Strakhovenko
\cite{baier} (see also \cite{Mitter,VillalbaChavez:2012bb}) were the first to determine  $\Pi_{\mu\nu}$   in the field  of a plane-wave  of the form
\begin{equation}
\mathscr{A}^{\mu}(x)=\mathpzc{a}_1^{\mu}\psi_1(\varkappa x)+\mathpzc{a}_2^{\mu}\psi_2(\varkappa x).\label{externalF}
\end{equation} Here $\mathpzc{a}_{1,2}$  are  the  amplitudes of the strong laser wave,  $\varkappa^{\mu}=(\varkappa^0,\pmb{\varkappa})$ denotes
its four-momentum while   $\psi_{1,2}$ are arbitrary  functions which characterize the shape of the laser field. The latter quantities additionally   fulfill
the following constraints:
\begin{wrapfigure}{r}{0.38\textwidth}
\includegraphics[width=.40\textwidth]{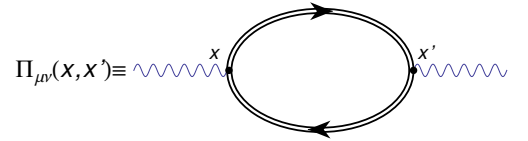}
\caption{\label{fig:mb000} Diagrammatical representation of the vacuum polarization tensor.  The double lines represent the electron-positron Green's functions
including the interaction with the external field. The two wavy lines denote the amputated legs corresponding
to the  small-amplitude electromagnetic waves.}
\end{wrapfigure}
\begin{equation}
\varkappa^2=0, \quad \varkappa \mathpzc{a}_1=\varkappa \mathpzc{a}_2=\mathpzc{a}_1\mathpzc{a}_2=0.\label{constraextf}
\end{equation}In this context, the external field tensor of  the wave  [Eq.~(\ref{externalF})] is $\mathscr{F}^{\mu\nu}=\sum_{i=1,2}\mathscr{F}^{\mu\nu}_i\psi_i^\prime(\varphi)$,
$\mathscr{F}^{\mu\nu}_{i}=\varkappa^\mu\mathpzc{a}^\nu_{i}-\varkappa^\nu\mathpzc{a}^\mu_{i}$,
  with $\varphi\equiv\varkappa x$ and $\psi_i^\prime(\varphi)\equiv d\psi_{i}/d \varphi$. It is worth noting at this point that  the constant electric
$[E^{j}_{i}=\mathscr{F}^{j0}_{i}]$ and magnetic $[\mathscr{F}^{jk}_{i}=-\epsilon^{jkl}B^{l}_{i}]$ amplitudes  associated with
each term in Eq.~(\ref{externalF}) are crossed, orthogonal  and with the same strength  $\vert\pmb{E}_{i}\vert=\vert\pmb{B}_{i}\vert.$

In the present paper the  field described above is also considered as the external background of the theory. Due to this fact  the  tensorial structure
of  $\Pi_{\mu\nu}$  can be written in terms of  Lorentz covariant vectors $\Lambda^\mu_{i}$  $(i=1,2,3,4)$:
\begin{eqnarray}\label{DVPBaier}
\Pi^{\mu\nu}(k_1,k_2)=c_1\Lambda^\mu_{1}\Lambda^\nu_{2}+c_2\Lambda^\mu_{2}\Lambda^\nu_{1}+c_3\Lambda^\mu_{1}\Lambda^\nu_{1}+c_4\Lambda^\mu_{2}\Lambda^\nu_{2}+c_5\Lambda^\mu_{3}\Lambda^\nu_{4}
\end{eqnarray}It is worth mentioning at this point that $\Lambda^\mu_{i}$  were constructed to satisfy the  first principles  of  charge conjugation,
spatial and time reversal symmetry as well as  gauge  and Poincar\'e  invariances in the polarization tensor.  Following the notation used in   \cite{baier}  we write
\begin{eqnarray}\label{vectorialbasisbaier}
\Lambda_{1}^\mu(k)=-\frac{\mathscr{F}_{1}^{\mu\nu}k_\nu}{(k\varkappa)\left(-\mathpzc{a}_1^2\right)^{\nicefrac{1}{2}}},\quad   \Lambda_{2}^\mu(k)=-\frac{\mathscr{F}_{2}^{\mu\nu}k_\nu}{(k\varkappa)\left(-\mathpzc{a}_2^2\right)^{\nicefrac{1}{2}}},\quad
\Lambda_{3}^\mu(k)=\frac{\varkappa^\mu k_1^2-k_{1}^{\mu} (k\varkappa)}{(k\varkappa)\left(k_1^2\right)^{\nicefrac{1}{2}}},\quad   \Lambda_{4}^\mu(k)=\frac{\varkappa^\mu k_2^2-k_{2}^{\mu} (k\varkappa)}{(k\varkappa)\left(k_2^2\right)^{\nicefrac{1}{2}}}.
\end{eqnarray} Note that the short-hand notation $k$ in the expressions above  may stand for either $k_1$ or $k_2.$
It is important to note  that the  vectors $\Lambda_1(k_1),$ $\Lambda_2(k_1)$ and $\Lambda_3(k_1)$   are orthogonal to  each other,   $\Lambda^\mu_{i}(k_1)\Lambda_{j\mu}(k_1)=-\delta_{ij}$,
and fulfill the completeness relation
\begin{equation}
\mathpzc{g}^{\mu\nu}-\frac{k_1^{\mu} k_1^{\nu}}{k_1^2}=-\sum_{i=1}^{3}\Lambda^\mu_{i}(k_1)\Lambda^\nu_{i}(k_1).\label{completness}
\end{equation}A similar statement applies if the set of vectors   $\Lambda_1(k_2),$ $\Lambda_2(k_2)$ and $\Lambda_4(k_2)$ are considered.

Now, in Eq.~(\ref{DVPBaier}) the form factor $c_i$  is  a distribution function which  depends on  the fundamental scalars of the theory:
$k^2$, $\varkappa k$ and $\xi_j^2=-e^2\mathpzc{a}_j^2/m_0^2$.  In order to determine  its low  energy behavior it is  convenient
to express the Fourier transformation of Eq.~(\ref{IRVPT}) as
\begin{eqnarray}\label{direcpimunu}
\Pi^{\mu\nu}(k_1,k_2)=&-&\mathfrak{L}_{\mathfrak{F}\mathfrak{F}}(\varkappa k_1)^2 \left\{\mathpzc{a}_1^2\Lambda^\mu_1\Lambda^\nu_1\int
\dbar^4 p\ \psi_1^\prime(p) \psi_1^\prime(k_1-k_2-p)-(-\mathpzc{a}^2_1)^{\nicefrac{1}{2}}(-\mathpzc{a}^2_2)^{\nicefrac{1}{2}}\left(\Lambda^\mu_1\Lambda^\nu_2+\Lambda^\mu_2\Lambda^\nu_1\right)
\right.\nonumber\\ &\times&\left.\int \dbar^4 p\ \psi_1^\prime(p) \psi_2^\prime(k_1-k_2-p)+\mathpzc{a}_2^2\Lambda^\mu_2\Lambda^\nu_2\int \dbar^4 p\ \psi_2^\prime(p) \psi_2^\prime(k_1-k_2-p)\right\}
-\mathfrak{L}_{\mathfrak{G}\mathfrak{G}}(\varkappa k_1)^2\left\{\mathpzc{a}_1^2\tilde{\Lambda}^\mu_1\tilde{\Lambda}^\mu_1 \nonumber\right.\nonumber\\
&\times&\left.\int\dbar^4 p\ \psi_1^\prime(p) \psi_1^\prime(k_1-k_2-p)-(-\mathpzc{a}^2_1)^{\nicefrac{1}{2}}(-\mathpzc{a}^2_2)^{\nicefrac{1}{2}}\left(\tilde{\Lambda}^\mu_1\tilde{\Lambda}^\nu_2
+\tilde{\Lambda}^\mu_2\tilde{\Lambda}^\nu_1\right)\int \dbar^4 p\ \psi_1^\prime(p)\psi_2^\prime(k_1-k_2-p)\right.\nonumber\\
&+&\left.\mathpzc{a}_2^2\tilde{\Lambda}^\mu_2\tilde{\Lambda}^\nu_2\int \dbar^4 p\ \psi_2^\prime(p) \psi_2^\prime(k_1-k_2-p)\right\},
\end{eqnarray} where $\psi_i^\prime(q)$ must be understood as  the Fourier transform of $\psi_i^\prime(\varphi)=d\psi_i(\varphi)/d\varphi$.
Note that the shorthand notation $\dbar^4 p\equiv d^4p/(2\pi)^4$ as well as  the  two  pseudovectors
\begin{eqnarray}\label{pseudovectors}
\tilde{\Lambda}_{1}^\mu(k)=-\frac{\tilde{\mathscr{F}}_{1}^{\mu\nu}k_\nu}{(k\varkappa)\left(-\mathpzc{a}_1^2\right)^{\nicefrac{1}{2}}},\quad
\tilde{\Lambda}_{2}^\mu(k)=-\frac{\tilde{\mathscr{F}}_2^{\mu\nu}k_\nu}{(k\varkappa)\left(-\mathpzc{a}_2^2\right)^{\nicefrac{1}{2}}}
\end{eqnarray} have been introduced. They are orthonormalized according to $\tilde{\Lambda}_i\tilde{\Lambda}_j=-\delta_{ij}$ and satisfy the relations
\begin{eqnarray}
\tilde{\Lambda}_i\Lambda_j=-\epsilon_{ij}, \quad \tilde{\Lambda}_i\Lambda_3=0 \quad  \mathrm{for} \quad i,j=1,2,\label{identity2}
\end{eqnarray} with the antisymmetric tensor $\epsilon_{ij}$   taken as $\epsilon_{12}=-\epsilon_{21}=1.$  We then  project $\Pi_{\mu\nu}(k_1,k_2)$ with
the appropriate combinations of the vectors $\Lambda_1\ldots\Lambda_4$  appearing in Eq.~(\ref{DVPBaier}). Guided by this procedure we find that
\begin{eqnarray}\label{formfactors1}
&&c_1=c_2,\quad c_3=c_4(1 \leftrightarrow 2),\quad c_5=0,\quad
c_2=-\frac{1}{15}\frac{\alpha}{\pi}\frac{(k\varkappa)^2}{m_0^2}\xi_1\xi_2\int \dbar^4 q\psi_1^\prime(q)\psi_2^\prime(k_1-k_2-q),\\
&&c_4=\frac{4}{45}\frac{\alpha}{\pi}\frac{(k\varkappa)^2}{m_0^2}\xi_2^2\int \dbar^4 q\psi_2^\prime(q)\psi_2^\prime(k_1-k_2-q)+\frac{7}{45}\frac{\alpha}{\pi}\frac{(k\varkappa)^2}{m_0^2}\xi_1^2\int \dbar^4 q\psi_1^\prime(q)\psi_1^\prime(k_1-k_2-q).\label{formfactors2}
\end{eqnarray} We want to stress  that the derivation of the these coefficients requires the use  of the  orthogonal character of the four-vectors  $\Lambda_i$,
as well as  Eq.~(\ref{identity2}).

The previous results show  that   the final structure of the vacuum polarization tensor in the field of a plane-wave [Eq.~(\ref{externalF})] depends on
its  specific shape.  This   statement  is manifest  through the absence of the usual Dirac  delta functions which impose   energy and
momentum conservation.  Therefore the interaction with a strong laser field  could, in general, involve  inelastic scattering.

\subsection{Equations of motion and general considerations in the case of an circularly polarized monochromatic wave}

Let us turn our attention to the equations of motion associated with our problem. These can be derived from the  Lagrangian [Eq.~(\ref{EASCPP})]
and turn out to be  coupled to each other. Indeed, in Landau  gauge $\partial_\mu a^\mu=0$  they read
\begin{eqnarray}\label{apepreliminar}
&&\left(\square+m^2\right)\phi-\frac{g}{8\pi}\tilde{\mathscr{F}}_{\mu\nu}f^{\mu\nu}=0,\\
&&\square a_\mu(x)+\int d^4x^\prime\Pi_{\mu\nu}(x,x^\prime)a^{\nu}(x^\prime)+g\tilde{\mathscr{F}}_{\mu\nu}\partial^\nu\phi=0. \label{paepreliminar}
\end{eqnarray}The first equation shows that  a small-amplitude electromagnetic wave can be converted into an axion via the
corresponding coupling  through the dual of the external field tensor. The second  equation, however,   allows   a reconversion
process in which the  axion  becomes  a propagating photon again.   In order to analyze such processes  it is  convenient to
transform into  momentum space. In this context,  Eqs.~(\ref{apepreliminar}) and  (\ref{paepreliminar}) are  coverted into  algebraic
forms
\begin{eqnarray}\label{photon-axion}
&&0=\Delta^{-1}(k)\phi(k)+\frac{ig}{4\pi} \sum_{i=1,2} \tilde{\Lambda}^\mu_i\varrho_i^{\nicefrac{1}{2}}\int \dbar^4 p \psi_i^\prime(p)a_\mu(k-p), \\
&&0=k^2a^{\mu}(k)-\int \dbar^4 q\  \Pi^{\mu\nu}(k,q)a_{\nu}(q)+ig\sum_{i=1,2}\tilde{\Lambda}^\mu_i\varrho_i^{\nicefrac{1}{2}}\int \dbar^4 p \psi_i^\prime(p)\phi(k-p)
\label{axion-photon}
\end{eqnarray} where $\varrho_i\equiv k_1\mathscr{F}_i^{2}k_1=k_1\tilde{\mathscr{F}}_i^{2}k_1$ is a Lorentz scalar  whose explicit structure  reads
\begin{eqnarray} \label{lorentzscalar}
\varrho_i= -\left(k\varkappa\right)^2\mathpzc{a}_i^2
\end{eqnarray}and  $\Delta^{-1}(k)=k^2-m^2$ is the inverse axion propagator. Note that  the  axion-photon  coupling is provided by the  pseudovectors  $\tilde{\Lambda}_{1,2}$  which   preserve
 parity invariance. Conversely, if the minimal coupling in Eq.~(\ref{actionpreliminar})  is replaced by a parity preserving  interaction
involving a scalar ALP,  i.e., $\sim g \phi \mathfrak{F},$   the mixing  term in Eq.~(\ref{EASCPP}) acquires a  structure  $\sim\frac{g}{8\pi} \mathscr{F}^{\mu\nu}f_{\mu\nu}.$
Following a procedure similar to that used  in this section  we obtain a system of  equations  similar to those given in Eqs.~(\ref{photon-axion}-\ref{axion-photon}),
the only difference arising in  the last term,  which now involves $\Lambda_i$  instead of $\tilde{\Lambda}_i$.

Let us consider the case in which the parameters $\xi_{1,2}$ and the   functions $\psi_{1,2}$ are chosen so that the external laser field  is  an circularly polarized monochromatic wave.
In our framework this corresponds to take $\xi^2\equiv\xi_1^2=\xi_2^2$,   $\psi_1=\cos\left(\varphi\right)$  and  $\psi_2=\sin\left(\varphi\right)$ with
\begin{eqnarray}\label{mpwft}
\psi_1^\prime(p)=\frac{1}{2i}\left[\deltabar^{(4)}\left(p+\varkappa\right)-\deltabar^{(4)}\left(p-\varkappa\right)\right],\quad
\displaystyle
\psi_2^\prime(p)=\frac{1}{2}\left[\deltabar^{(4)}\left(p+\varkappa\right)+\deltabar^{(4)}\left(p-\varkappa\right)\right],
\end{eqnarray} and \     $\deltabar^{(4)}(x)\equiv(2\pi)^4\delta^{(4)}\left(x\right)$. This particular context  allows for introducing  the following covariant vectors:
\begin{equation}\label{solidaridad}
\Lambda^\mu_{\pm}=\Lambda^\mu_1\pm i\Lambda^\mu_2\quad \mathrm{and}\quad\tilde{\Lambda}^\mu_{\pm}=\tilde{\Lambda}^\mu_1\pm i\tilde{\Lambda}^\mu_2,
\end{equation} where  $\Lambda_{1,2}$  are given in Eq.~(\ref{vectorialbasisbaier}). Note that the new vectors in Eq.~(\ref{solidaridad}) satisfy the  relations
\begin{eqnarray}
\begin{array}{c}
\Lambda_+\Lambda_-=-2,\ \ \tilde{\Lambda}_+\Lambda_-=-\tilde{\Lambda}_-\Lambda_+=2i,\quad
\Lambda_+\Lambda_+=\Lambda_-\Lambda_-=\Lambda_+\tilde{\Lambda}_+=\Lambda_-\tilde{\Lambda}_-=0.
\end{array}
\end{eqnarray} At this point,  it is worth noting  that the  low energy behavior of the vacuum polarization tensor can be written as
\begin{eqnarray}\label{circualrformfactor}
\begin{array}{c}\displaystyle
\Pi^{\mu\nu}(k_1,k_2)=\sum_{n=0,+,-}\Pi^{\mu\nu}_{n}\deltabar^{(4)}\left(k_1-k_2+2n\varkappa\right),\\  \displaystyle
\Pi^{\mu\nu}_0=\pi_3(\Lambda^\mu_1\Lambda^\nu_1+\Lambda^\mu_2\Lambda^\nu_2),\quad
\Pi^{\mu\nu}_{\pm}=\pi_0 \Lambda^\mu_{\pm}\Lambda^\nu_{\pm},\quad \pi_3=\frac{11}{90}\frac{\alpha}{\pi}\frac{(\varkappa k)^2}{m_0^2}\xi^2,\qquad \pi_0=\frac{1}{60}\frac{\alpha}{\pi}\frac{(\varkappa k)^2}{m_0^2}\xi^2.
\end{array}
\end{eqnarray}  The structure of these entities coincides with those obtained by Ba\u{\i}er, Mil'shte\u{\i}n and Strakhovenko in  \cite{baier} -
according to the correspondence $\pi_3\Leftrightarrow\alpha_3$ and   $\pi_0\Leftrightarrow\alpha_0$. Observe that the tensorial structures  $\Pi^{\mu\nu}_{n}$
are in correspondence with the possible states  of helicity  $n=0,+,-$. 

Eq.~(\ref{circualrformfactor}) warrants  further comment. Firstly, it may be seen that the  scattered field is  emitted  with three different
frequencies. One of these   coincides with the frequency of the  incoming  small-amplitude wave, resulting in an  elastic scattering. The remaining two frequencies
emerge as a consequence of  inelastic processes  in  which the emission and absorption of two laser  photons occur. These turn out to be shifted to lower
and  higher values in comparison with  the original monochromatic frequency. The scattering of light in these latter two cases is analogous to the Raman process
in molecular physics  with $\varkappa_0$ imitating the vibrational frequency of the molecules.  Similarly, it  might be used to  test the nonlinear properties  of  the
QED vacuum. In fact, the associated spectroscopy has been recently put forward as alternative way of probing the predicted
vacuum of minicharged particles \cite{Villalba-Chavez:2013txu}.

In  order to pursue our research  we  insert  Eqs.~(\ref{mpwft}) and  (\ref{circualrformfactor}) into Eqs.~(\ref{photon-axion}) and (\ref{axion-photon}), arriving at  the following
equation for the photon-axion
\begin{eqnarray}
\Delta^{-1}(k)\phi(k)-\frac{g}{8\pi}\varrho^{\nicefrac{1}{2}}\tilde{\Lambda}^{\mu}_{-}a_\mu(k-\varkappa)+\frac{g}{8\pi}\varrho^{\nicefrac{1}{2}}\tilde{\Lambda}^{\mu}_{+}a_\mu(k+\varkappa)=0,
\label{photon-axion-circular1}
\end{eqnarray}
and axion-photon conversion
\begin{eqnarray}
k^2a^{\mu}(k)+\frac{1}{2}g\varrho^{\nicefrac{1}{2}}\left[\tilde{\Lambda}^\mu_+\phi(k+\varkappa)-\tilde{\Lambda}^\mu_-\phi(k-\varkappa)\right]-\sum_{\lambda=0,+,-}\Pi^{\mu\nu}_{\lambda}(k)a_{\nu}(k+2\lambda\varkappa)=0,\label{axion-photon-circular}
\end{eqnarray}where $\varrho\equiv \varrho_1=\varrho_2$ is the  Lorentz scalar given in Eq.~(\ref{lorentzscalar}).
These  last two equations   constitute  our starting point for the following  analyses. They  reveal  that the conversion process  changes the momentum content.
Thus, in presence of a circularly  polarized monochromatic wave  the mixing phenomenon is conceptually
more involved than in the case of a constant magnetic field.

The  solution of our problem  can be written as a superposition  of two transverse  waves
\begin{equation}\label{chiralityexpansio}
a^{\mu}(k)=\frac{f_+(k)}{\sqrt{2}}\Lambda^\mu_++\frac{f_-(k)}{\sqrt{2}}\Lambda^\mu_-.
\end{equation}Two additional terms may be included  in this expansion. However, both are associated with longitudinal and nonphysical propagation
modes. One of these terms is  longitudinal by construction $\sim k^\mu$; while the remaining is transverse  and proportional to  $\Lambda^\mu_3$,
the  absence of $c_5$ in Eq.~(\ref{circualrformfactor}) [compare with Eq.~(\ref{DVPBaier})] leads to a trivial dispersion equation $k^2=0$ and so
$\Lambda^\mu_3\sim k_\mu$  [see  Eq.~(\ref{vectorialbasisbaier})] becomes a longitudinal gauge  mode.  As such, both solutions have been  omitted.

We  substitute  Eq.~(\ref{chiralityexpansio}) into Eqs.~(\ref{photon-axion-circular1})-(\ref{axion-photon-circular})  and  multiply $\Lambda_{\pm}^\mu$ by  the  left-hand side of
Eq.~(\ref{axion-photon-circular}). As a consequence,   the resulting system of equations to be analyzed is
\begin{equation}\label{eigenproblem}
\pmb{\mathscr{G}}^{(i)}(k)\pmb{z}^{(i)}(k)=0\quad \mathrm{with}\quad i=1,2.
\end{equation}Here the  quantities involved are defined as follows
\begin{equation}
\pmb{\mathscr{G}}^{(1)}(k)=\left[
\begin{array}{ccc}
\Delta^{-1}(k+\varkappa)& \frac{i\sqrt{2}}{8\pi}g\varrho^{\nicefrac{1}{2}}& \frac{i\sqrt{2}}{8\pi}g\varrho^{\nicefrac{1}{2}}\\
-\frac{i\sqrt{2}}{2}g\varrho^{\nicefrac{1}{2}} & k^2+\pi_3 &2\pi_0\\
-\frac{i\sqrt{2}}{2}g\varrho^{\nicefrac{1}{2}} & 2\pi_0 & (k+2\varkappa)^2+\pi_3
\end{array}\right],\quad
\pmb{\mathscr{G}}^{(2)}(k)=\left[
\begin{array}{ccc}
\Delta^{-1}(k-\varkappa)& \frac{i\sqrt{2}}{8\pi}g\varrho^{\nicefrac{1}{2}}& \frac{i\sqrt{2}}{8\pi}g\varrho^{\nicefrac{1}{2}}\\
-\frac{i\sqrt{2}}{2}g\varrho^{\nicefrac{1}{2}} & (k-2\varkappa)^2+\pi_3 &2\pi_0\\
-\frac{i\sqrt{2}}{2}g\varrho^{\nicefrac{1}{2}} & 2\pi_0 & k^2+\pi_3
\end{array}\right],
\end{equation}
\begin{equation}
\pmb{z}^{(1)} =
\left[\begin{array}{c}
\phi(k+\varkappa)\\ f_+(k)\\ f_-(k+2\varkappa)
\end{array}\right],\quad \pmb{z}^{(2)} =
\left[\begin{array}{c}
\phi(k-\varkappa)\\ f_+(k-2\varkappa)\\ f_-(k)
\end{array}\right].\label{statesflavornomass}
\end{equation}It is remarkable that  both  eigenproblems are correlated by means of the relations
\[
\pmb{\mathscr{G}}^{(1)}(k-2\varkappa)\pmb{z}^{(1)}( k-2\varkappa)=\pmb{\mathscr{G}}^{(2)}(k)\pmb{z}^{(2)}(k)=0\quad \mathrm{and} \quad\pmb{\mathscr{G}}^{(2)}(k+2\varkappa)\pmb{z}^{(2)}( k+2\varkappa)=\pmb{\mathscr{G}}^{(1)}(k)\pmb{z}^{(1)}(k)=0.
\]
We point out that  the field components   contained in these  vectors  cannot be understood as  mass eigenmodes. Once
the ALP-photon coupling is considered they become--as occurs in the neutrino oscillations \cite{Kuo:1989qe}--``flavor'' eigenstates. This means that
the fields in the Lagrangian are not equivalent to the mass eigenstates/propagating modes of the interacting theory.

\section{Oscillations \label{circulaexternafield}}

\subsection{Isolating the ALP-induced vacuum birefringence \label{DR}}

Nontrivial solutions of  the mixing process emerge whenever  the determinant of $\pmb{\mathscr{G}}^{(1)}(k)$   vanishes identically.  In such a case, a cubic equation in $k^2$ (sextic
in the frequency $\mathpzc{w}$) is generated:
\begin{eqnarray}\label{eqdtrab2}
&&\left(k^2+\pi_3\right)\left[k^2-m^2
+2 \left(k \varkappa\right)\right]\left[(k+2\varkappa)^2
+\pi_3\right]=\frac{g^2}{4\pi} \varrho \left[k^2+2\left(k\varkappa\right)\right],
\end{eqnarray} where the dispersion equation for the strong wave, i.e., $\varkappa^2=0$ has been used. Moreover,  this outcome  has been derived by neglecting those  terms  resulting
from the off-diagonal components of $\pmb{\mathscr{G}}^{(1)}(k)$ which are proportional to $\sim\alpha^2$ and $\sim g^2\alpha$.
Note that   the left-hand side  of this equation still contain contributions that will be  eventually disregarded  as, e.g., a term proportional to  $\sim\pi_3^2$. Certainly,
the  exact solutions of Eq.~(\ref{eqdtrab2}) can be determined by analytical procedures.
However, we are interested in  analyzing  the physical context  in which  the  ALP-photon coupling  does not  dramatically modify  the free  dispersion relations of the  particles
involved. Accordingly, one can manipulate the right-hand side in Eq.~(\ref{eqdtrab2})  as a small perturbative  correction to the leading  equations
 which  result  when the ALP-photon coupling vanishes identically. This assumption allows us  to  apply a recursive method where the following
set of  equations
\begin{eqnarray}
&&k^2=0,\label{leadingdispeqsmassless}\\
&&k^2-m^2+2 \left(k \varkappa\right)=0 \label{leadingdispeqsmassive}
\end{eqnarray} is taken as the  starting point.  In this framework, two   massless modes are found\footnote{This result is somewhat expected, starting with two photon modes, a massive axion mode and assuming a tiny coupling/mixing.}. The first  is  determined  by passing  the square
brackets in Eq.~(\ref{eqdtrab2})  to the  right-hand side.  We  use Eq.~(\ref{leadingdispeqsmassless})  to express  the resulting equation as  follows
\begin{eqnarray}\label{intermediate}
k^2\simeq-\pi_3+\frac{g^2\varrho}{8\pi(2k\varkappa-m^2)}.
\end{eqnarray} The left-hand side of this expression is  then linearized   with respect to $\mathpzc{w}$ by approaching
$k^2=\mathpzc{w}^2-\pmb{k}^2=(\mathpzc{w}-\vert\pmb{k}\vert)(\mathpzc{w}+\vert\pmb{k}\vert)\vert\approx2\vert\pmb{k}\vert(\mathpzc{w}-\vert\pmb{k}\vert)$.
We additionally set, the  momentum $k$  involved on   the right-hand side of this equation  to $k=\mathpzc{k}\equiv(\omega_{\pmb{k}},\pmb{k})$ with
$\omega_{\pmb{k}}\equiv\vert\pmb{k}\vert.$  As a consequence the dispersion relation is found to be
\begin{eqnarray}\label{photonlikesolution}
\mathpzc{w}_+^{(1)}(\pmb{k})\approx\omega_{\pmb{k}} -\frac{\pi_3}{2\omega_{\pmb{k}}} +\frac{g^2 (\mathpzc{k} \varkappa)^2 I }{4\varkappa_0^2\omega_{\pmb{k}}\left[2(\mathpzc{k} \varkappa)-m^2\right]},
\end{eqnarray}where  $I=E^2/4\pi=\varkappa_0^2\pmb{\mathpzc{a}}^2/(4\pi)$ denotes  the peak intensity  associated with  strong  field of the wave\footnote{Observe that the  temporal gauge, i.e., $\mathpzc{a}_0=0$ has been chosen.}. The  second massless solution
can be determined by  moving  the first two  brackets in Eq.~(\ref{eqdtrab2})  to its right-hand side and setting $k= \mathpzc{k}+2\varkappa$. We then use  the linearization
\begin{equation}\label{linerraman}
(k+2\varkappa)^2\simeq2\omega_{\pmb{k}+2\pmb{\varkappa}}\left(\mathpzc{w}-\omega_{\pmb{k}+2\pmb{\varkappa}}+2\varkappa_0\right),
\end{equation}which applies for  $k\varkappa\simeq0$. As a consequence, it follows
\begin{eqnarray}
\mathpzc{w}_{-}^{(1)}(\pmb{k}+2\pmb{\varkappa})=\omega_{\pmb{k}+2\pmb{\varkappa}}-2\varkappa_0-\frac{\pi_3}{2\omega_{\pmb{k}+2\pmb{\varkappa}}}-\frac{g^2 (\mathpzc{k} \varkappa)^2 I}{4 \varkappa_0^2\omega_{\pmb{k}+2\pmb{\varkappa}}\left[2(\mathpzc{k} \varkappa)+m^2\right]},
\label{dispersionrelationmode2}
\end{eqnarray}where the short-hand notation $\omega_{\pmb{k}+2\pmb{\varkappa}}\equiv\vert\pmb{k}+2\pmb{\varkappa}\vert$ has been introduced. Note that the subindices of  $\mathpzc{w}_\pm^{(1)}$
have been added to establish a correspondence between the dispersion relations  and  the   helicity states.

Some comments are in order. Firstly  in the limit where   $\omega_{\pmb{k}}\to0$ the dispersion relations  [Eqs.~(\ref{photonlikesolution}) and  (\ref{dispersionrelationmode2})]
become trivial. As such  gauge invariance is preserved and one can  identify Eqs.~(\ref{photonlikesolution})  and   (\ref{dispersionrelationmode2}) as the
photon-like solutions of the mixing process. The pole in the interacting  term of  $\mathpzc{w}_+^{(1)}(\pmb{k})$  also deserves some attention.  This  translates  into an ALP mass depending
not only upon  the  momentum of the probe laser beam  but also on the frequency of the strong background field
\begin{equation}\label{resonancemass}
m_{*}=\left(2\mathpzc{k}\varkappa\right)^{\nicefrac{1}{2}}.
\end{equation}When the above condition  is fulfilled the dispersion relation Eq.~(\ref{photonlikesolution})  is resonantly enhanced. Obviously, this is  not consistent with our perturbative treatment.
However,  Eq.~(\ref{photonlikesolution})  can be  used to explore the domains in which the   ALP mass is  near  resonance, i.e.,
$m=m_\mathrm{*}\pm\epsilon,$ $\epsilon>0$ provided the condition
\begin{equation}
m_\mathrm{*}\gg\epsilon\gg \frac{g^2m_*^3 I}{32\varkappa_0^2\omega_{\pmb{k}}^2 }. \label{closeresonace}
\end{equation} Otherwise the use  of our perturbative approach would  not be  justified.
Nevertheless, whenever the  collision angle between the two waves  is tiny  [$\theta\ll1$],
small    resonant masses might   be  explored:
\begin{equation}\label{smallanhelrr}
m_*\simeq \theta\left(\omega_{\pmb{k}}\varkappa_0\right)^{\nicefrac{1}{2}}.
\end{equation}However, less stringent constraints in the coupling constant $g$ are  expected to appear because the interaction becomes
extremely small $\sim\theta^3$. This case will be  treated  shortly.

Our original dispersion equation [Eq.~(\ref{eqdtrab2})] also allows for   massive solutions. In order to determine which of them are  physical,  we first note
that  Eq.~(\ref{leadingdispeqsmassive}) provides two frequencies.  We discard  the one  which  is negative  when the
external field frequency tends to zero. In correspondence, we obtain
\begin{eqnarray}
\begin{array}{c}
\omega_+=\varepsilon_{\pmb{k}+\pmb{\varkappa}}-\varkappa_0,\quad
\varepsilon_{\pmb{k}+\pmb{\varkappa}}=\left[(\omega_{\pmb{k}}+\varkappa_0)^2+m^2-m_*^2\right]^{\nicefrac{1}{2}}.
\end{array}
\label{celtic}
\end{eqnarray} This expression is  in agreement with the energy-momentum conservation whose balance at tree level reads
\[
k+\varkappa=p_+\quad\mathrm{with}\quad p_+^\mu=(\varepsilon_{\pmb{p}},\pmb{p}).
\]As long  as Eq.~(\ref{leadingdispeqsmassless}) is taken into account, the above relation promotes the resonant condition $m^2=m_*^2$.  Observe that,  in  the vicinity of the  resonance,
$m^2-m_*^2\approx2m_*\epsilon$ and in correspondence Eq.~(\ref{celtic}) can be written as
\begin{eqnarray}
\omega_+\simeq\omega_{\pmb{k}}+\frac{1}{2}\frac{m^2- m_*^2}{\omega_{\pmb{k}}+\varkappa_0},\label{treelevelsolutionmassivemodes}
\end{eqnarray}where the second term must be understood as a very small  contribution with $(\omega_{\pmb{k}}+\varkappa_0)^2\gg 2 \epsilon m_*$. 
The  correction  to Eq.~(\ref{treelevelsolutionmassivemodes}) due to the ALP-photon coupling can be found  similarly  to how  the massless modes were determined.
Using the  linearization $k^2+m_*^2-m^2\simeq2\varepsilon_{\pmb{k}+\pmb{\varkappa}}(\mathpzc{w}-\omega_+)$, we find that the massive solution of Eq.~(\ref{eqdtrab2})--up to  first
nontrivial order in $g^2$--is given by
\begin{equation}\label{posmassmod}
\mathpzc{w}_0^{(\mathrm{1})}(\pmb{k})\approx\omega_{\pmb{k}}+\frac{1}{2}\frac{m^2-m_*^2}{\omega_{\pmb{k}}+\varkappa_0}+\frac{g^2 I m^2 m_*^4}{8\varkappa_0^2\varepsilon_{\pmb{k}+\pmb{\varkappa}}\left[m^4-m_*^4\right]}.
\end{equation} The above expression   diverges when the mass coincides with the resonant one
[Eq.~(\ref{resonancemass})].  However, similar to the massless mode [Eq.~(\ref{dispersionrelationmode2})], it can be exploited to investigate the ALP-photon oscillations
near resonance, provided  Eq.~(\ref{closeresonace}) is satisfied.

In order to  determine the solutions of the  remaining eigenproblem, the  determinant of $\pmb{\mathscr{G}}^{(2)}$  must vanish. This condition  generates  the   dispersion equation
\begin{eqnarray}\label{decp2}
\left(k^2+\pi_3\right)\left[k^2-m^2
-2 \left(k \varkappa\right)\right]\left[(k-2\varkappa)^2
+\pi_3\right]=\frac{g^2}{4\pi} \varrho \left[k^2-2\left(k\varkappa\right)\right].
\end{eqnarray} The recursive  procedure described  above allows us to  find  a  photon-like solution associated with the negative helicity mode
\begin{equation}\label{cirmodes1}
\mathpzc{w}_-^{(2)}(\pmb{k})\approx \omega_{\pmb{k}}-\frac{\pi_3}{2\omega_{\pmb{k}}} -\frac{g^2 m_*^4 I }{16\varkappa_0^2\omega_{\pmb{k}}\left[m_*^2+m^2\right]}.
\end{equation}In contrast, the dispersion law for a  photon-like state with  positive helicity and momentum $\pmb{k}-2\pmb{\varkappa}$ reads
\begin{eqnarray}\label{cirmodes2}
\mathpzc{w}_{+}^{(2)}(\pmb{k}-2\pmb{\varkappa})\approx\omega_{\pmb{k}-2\pmb{\varkappa}}+2\varkappa_0-\frac{\pi_3}{2\omega_{\pmb{k}-2\pmb{\varkappa}}}
+\frac{g^2m_*^4 I}{16 \varkappa_0^2\omega_{\pmb{k}-2\pmb{\varkappa}}\left[m_*^2-m^2\right]},
\end{eqnarray}where $ \omega_{\pmb{k}-2\pmb{\varkappa}}=\vert\pmb{k}-2\pmb{\varkappa}\vert$. Clearly, another massive solution arises from Eq.~(\ref{decp2}).
The starting point for finding out this dispersion law  is the leading order equation $k^2-2k\varkappa-m^2=0$. Among its solutions, the following becomes
noticeable
\begin{eqnarray}
\begin{array}{c}
\omega_-=\varepsilon_{\pmb{k}-\pmb{\varkappa}}+\varkappa_0, \quad \varepsilon_{\pmb{k}-\pmb{\varkappa}}=\left[(\omega_{\pmb{k}}-\varkappa_0)^2+m_*^2+m^2\right]^{\nicefrac{1}{2}}.
\end{array}
\label{treelevelsoldutionmassivemodes}
\end{eqnarray} This  describes the energy conservation of  a   hypothetical  mixing  where the probe beam emits  a photon of the strong wave.
This kind  of oscillations  are  kinematically forbidden at  tree level  since the energy-momentum balance $k-\varkappa=p_-$ with $p_-^\mu=(\varepsilon_{\pmb{p}},\pmb{p})$
implies a  process where the ALP mass is  negative $m^2=-m_*^2\leqslant0$.  Once  the corrections coming from  the vacuum polarization and the ALP-photon
interaction are incorporated,  the dispersion equation [Eq.~(\ref{decp2})]  replaces the previous  condition and another massive solution could  arise. In
Sec.~\ref{conversionprosect} we will show that the nonoccurrence of the aforementioned  process--at tree level--is  intrinsically associated with  the monochromaticity
of  the  strong   wave [Eq.~(\ref{externalF})], a fact which formally restricts us to work in the limit of infinite pulse length.  Nevertheless, in practice  the interaction
time is always finite, the strong wave is not monochromatic and,  consequently, one can approach the remaining massive solution by
\begin{equation}\label{negmassmod}
\mathpzc{w}_0^{(2)}(\pmb{k})\approx\omega_{\pmb{k}}+\frac{1}{2}\frac{m^2+m_*^2}{\omega_{\pmb{k}}-\varkappa_0}+\frac{g^2 I m^2 m_*^4}{8 \varkappa_0^2\varepsilon_{\pmb{k}-\pmb{\varkappa}}\left[m^4-m_*^4\right]},
\end{equation}where the approximation $(\omega_{\pmb{k}}-\varkappa_0)^2\gg m^2+m_*^2$ has been used.  Clearly, in the limit of $g\to0$ the
above dispersion relation reduces to a nonphysical tree level condition which is connected to   Eq.~(\ref{treelevelsoldutionmassivemodes}). However, we will see very shortly
that in such a context,  a vanishing probability of conversion is obtained. Moreover, it will be shown  that, as soon  as the ALP-photon interaction is taken into account,
the probability that a probe photon oscillates into $\phi(k-\varkappa)$  is very small in comparison with the remaining possibility of mixing, i.e., when  $\phi(k+\varkappa)$ is involved.
This situation is  somewhat expected:  among the massive-like solutions $\mathpzc{w}_0^{(1)}(\pmb{k})$ defines the state with minimal energy.
Hence, the conversion of a  photon into a massive mode with energy $\mathpzc{w}_0^{(2)}$ is less likely to occur.

To  conclude this subsection we determine the phase velocity $\mathpzc{v}_\pm=\mathpzc{w}_\pm(\pmb{k})/\vert\pmb{k}\vert$ associated with each massless propagation mode,
i.e.,  Eqs.~(\ref{photonlikesolution}) and (\ref{cirmodes1}). In this case  we find
\begin{equation} \label{phasevelocitycircular}
\mathpzc{v}_\pm=1-\frac{\pi_3}{2\omega_{\pmb{k}}^2}\pm\frac{g^2m_*^4 I}{16\varkappa_0^2\omega_{\pmb{k}}^2\left(m_*^2\mp m^2\right)}.
\end{equation}
Obviously,  in the  absence of the ALP-photon coupling, both modes propagate with the same   phase velocity
$\mathpzc{v}_\pm\simeq1-\pi_3/2\omega_{\pmb{k}}^2.$ This implies that, at lower energy-momentum transfer [$\omega_{\pmb{k}}$, $\varkappa_0\ll m_0$]
and  in the weak field approximation [$E\ll E_c$], the QED vacuum in the field of a circular polarized wave--in leading order--behaves as an isotropic nonbirefringent medium. This  situation, however,
is  reverted  when the ALP-photon   coupling is considered.  In fact, the last  term in Eq.~(\ref{phasevelocitycircular}) manifests  that  the  plausible  emission and  absorption
of virtual ALPs with different momentum content  induces  a  chiral-like  birefringence.

\subsection{The flavor-like states \label{secflavorstates}}

The  previous linearizations in the dispersion equations  are equivalents to  reduce  the differential order  in the equations
of motion [Eqs.~(\ref{apepreliminar})-(\ref{paepreliminar})]. In correspondence,  we can approach  the first flavor-like state in Eq.~(\ref{statesflavornomass})  as
a superposition of the three mass eigenstates  which   characterize the mixing process
\begin{equation}\label{phasespacesolution}
\pmb{z}^{(1)}(\omega)\simeq\sum_{\lambda=0,+,-}\mathscr{N}_\lambda^{(1)} \pmb{z}_{\lambda}^{(1)}\delta\left(\omega-\mathpzc{w}_\lambda^{(1)}\right).
\end{equation}While $\mathscr{N}_\lambda^{(1)}$ denote some constants to be determined by the initial conditions, $\pmb{z}_\lambda^{(1)}$ represent the normalized eigenstates
of $\pmb{\mathscr{G}}^{(1)}$:
\begin{eqnarray}\label{normalizeeigenstates1}
\pmb{z}_{+}^{(1)}=\frac{\left[i\tan\left(\theta_{+}^{(1)}\right),  1, -\tan\left(\varphi_+^{(1)}\right)\right]}{\left[1+\tan^2\left(\theta_+^{(1)}\right)+\tan^2\left(\varphi_0^{(1)}\right)\right]^{\nicefrac{1}{2}}},\quad
\pmb{z}_{0}^{(1)}=\frac{\left[1,i\tan\left(\theta_0^{(1)}\right), i\tan\left(\varphi_0^{(1)}\right)\right]}{\left[1+\tan^2\left(\theta_0^{(1)}\right)+\tan^2\left(\varphi_0^{(1)}\right)\right]^{\nicefrac{1}{2}}},\quad
\pmb{z}_{-}^{(1)}=\frac{\left[i\tan\left(\theta_-^{(1)}\right), \tan\left(\varphi_-^{(1)}\right),1\right]}{\left[1+\tan^2\left(\theta_-^{(1)}\right)+\tan^2\left(\varphi_-^{(1)}\right)\right]^{\nicefrac{1}{2}}}.
\end{eqnarray} It is convenient to emphasize that these eigenstates have been calculated with accuracy of terms $\sim \mathpzc{o}(g^2),$ $\sim \mathpzc{o}(\alpha^2)$ and $\sim \mathpzc{o}(g \alpha)$.
Here, the pair $\theta_+^{(1)}, \theta_0^{(1)}$ parametrizes the ALP-photon oscillations in which photons with positive helicity are involved. Explicitly,
\begin{eqnarray}\label{mixingangleweakapp1}
&&\tan\left(\theta_+^{(1)}\right)=\left.\frac{\phi(k+\varkappa)}{if_+(k)}\right\vert_{\omega=\mathpzc{w}_+^{(1)}}=\frac{g m_*^2\sqrt{I}}{8\sqrt{2\pi}\varepsilon_{\pmb{k}+\pmb{\varkappa}}\varkappa_0\left(\varepsilon_{\pmb{k}+\pmb{\varkappa}}-\varkappa_0-\omega_{\pmb{k}}\right)},\\
&&\tan\left(\theta_0^{(1)}\right)=\left.\frac{f_+(k)}{i\phi(k+\varkappa)}\right\vert_{\omega=\mathpzc{w}_0^{(1)}}=\frac{g m_*^2\pi\sqrt{ I}}{2\sqrt{2\pi}\omega_{\pmb{k}}\varkappa_0\left(\varepsilon_{\pmb{k}+\pmb{\varkappa}}-\varkappa_0-\omega_{\pmb{k}}\right)},\\
&&\tan\left(\theta_-^{(1)}\right)=\left.\frac{\phi(k+\varkappa)}{if_-(k+2\varkappa)}\right\vert_{\omega=\mathpzc{w}_-^{(1)}}=\frac{gm_*^2\sqrt{I}}{8\sqrt{2\pi}\varepsilon_{\pmb{k}+\pmb{\varkappa}}\varkappa_0\left(\varepsilon_{\pmb{k}+\pmb{\varkappa}}+\varkappa_0-\omega_{\pmb{k}+2\pmb{\varkappa}}\right)}.\label{mixingandgleweakapp1}
\end{eqnarray} 
We stress that  $m_*$ is given in  Eq.~(\ref{resonancemass}). On the other hand, the expression of  $\varepsilon_{\pmb{k}+\pmb{\varkappa}}$  can be read off 
from Eq.~(\ref{celtic}) and  (\ref{treelevelsolutionmassivemodes}). The remaining angles contained in  $\pmb{z}_\lambda^{(1)}$  describe the  mixing between  photons with  different helicities. They read
\begin{eqnarray}
&&\tan\left(\varphi_+^{(1)}\right)=-\left.\frac{f_-(k+2\varkappa)}{f_+(k)}\right\vert_{\omega=\mathpzc{w}_+^{(1)}}
=\frac{\pi_0}{\omega_{\pmb{k}+2\pmb{\varkappa}}\left(\omega_{\pmb{k}}-\omega_{\pmb{k}+2\pmb{\varkappa}}+2\varkappa_0\right)}\label{photonmixing1},\\
&&\tan\left(\varphi_0^{(1)}\right)=-\left.\frac{f_-(k+2\varkappa)}{i\phi(k+\varkappa)}\right\vert_{\omega=\mathpzc{w}_0^{(1)}}=\frac{gm_*^2\pi \sqrt{I}}{2\sqrt{2\pi}\omega_{\pmb{k}+2\pmb{\varkappa}}\varkappa_0 \left(\omega_{\pmb{k}}-\omega_{\pmb{k}+2\pmb{\varkappa}}+2\varkappa_0\right)}\label{photonmixing2},\\
&&\tan\left(\varphi_-^{(1)}\right)=\left.\frac{f_+(k)}{f_-(k+2\varkappa)}\right\vert_{\omega=\mathpzc{w}_-^{(1)}}
=\frac{\pi_0}{\omega_{\pmb{k}}\left(\omega_{\pmb{k}}-\omega_{\pmb{k}+2\pmb{\varkappa}}+2\varkappa_0\right)},\label{photonmixing3}
\end{eqnarray} where the explicit expression of   $\pi_0$ can be found in  Eq.~(\ref{circualrformfactor}).

We continue our analysis by Fourier transforming  Eq.~(\ref{phasespacesolution}) only in time. Next,  we  consider  the experimental setup in  which the
incoming  probe beam is  a linearly  polarized plane wave. Upon entering  in the region occupied by  the external field of the wave, the probe beam is decomposed into
its circular-polarized waves  [Eq.~(\ref{chiralityexpansio})]. In connection,  we  suppose  that  at $t=0$  only the  incoming  beam has a nonvanishing amplitude with
$f_{\pm}(\pmb{k},0)=a_0$. Guided by this procedure, one  obtains  a system of algebraic equations for $\mathscr{N}_\lambda^{(1)}.$  Its solution allows us to express the
flavor-like components in the following form:
\begin{eqnarray}\label{afield}
\begin{array}{c}\displaystyle
f_+(\pmb{k},t)\simeq a_0e^{-i\mathpzc{w}_+^{(1)}t} \left\{1-\theta_+^{(1)}\theta_0^{(1)}\left[1-e^{i\left(\mathpzc{w}_+^{(1)}-\mathpzc{w}_0^{(1)}\right)t}\right]-\varphi_+^{(1)}\varphi_-^{(1)}\left[1-e^{i\left(\mathpzc{w}_+^{(1)}-\mathpzc{w}_-^{(1)}\right)t}\right]\right\},\\ \\
\displaystyle
\phi(\pmb{k}+\pmb{\varkappa},t)\simeq-ia_0\theta_+^{(1)}e^{-i\left(\mathpzc{w}_0^{(1)}+\varkappa_0\right)t}\left[1-e^{i\left(\mathpzc{w}_0^{(1)}-\mathpzc{w}_+^{(1)}\right)t}\right],\quad
f_-(\pmb{k}+2\pmb{\varkappa},t)\simeq a_0
 \varphi_+^{(1)} e^{-i\left(\mathpzc{w}_-^{(1)}+2\varkappa_0\right)t}\left[1-e^{i\left(\mathpzc{w}_-^{(1)}-\mathpzc{w}_+^{(1)}\right)t}\right],\label{photonmenos1}
\end{array}
 \end{eqnarray} where the  approximations of weak mixing [$\theta_\lambda^{(1)},\varphi_\lambda^{(1)}\ll1$]   have been used.  The  solutions found in this way
reveal that   the  outgoing probe  beam  contains  electromagnetic  radiation  resulting from  the  inelastic scattering. These kind of evanescent waves  should
emerge, in first instance,  due to the  vacuum polarization effects. Note that Eq.~(\ref{afield}) neither depend on  $\theta_-^{(1)}$ nor  $\varphi_0^{(1)}$.
This is because they are  associated with higher order  processes\footnote{For instance, the oscillations between the  Raman-like waves  and the axion  field.} whose contributions can be ignored.

The determination of  the flavor-like fields associated with the second eigenproblem  is quite similar to the  case previously  analyzed. Following  the same  line of reasoning,
we note   that the normalized eigenstates of $\pmb{\mathscr{G}}^{(2)}(k)$ can be found  from Eq.~(\ref{normalizeeigenstates1}), provided
the replacement  $1\to2$. The corresponding mixing angles can  be  obtained from Eqs.~(\ref{mixingangleweakapp1})-(\ref{photonmixing3}) by applying the symmetry transformation
that connects  both  eigenproblems [see below Eq.~(\ref{statesflavornomass})]. However,  in contrast to the previous case, the leading order terms of  the flavor-like
 fields are given by
\begin{eqnarray}
\begin{array}{c}\displaystyle
f_-(\pmb{k},t)\simeq a_0e^{-i\mathpzc{w}_-^{(2)}t} \left\{1-\theta_-^{(2)}\theta_0^{(2)}\left[1-e^{i\left(\mathpzc{w}_-^{(2)}-\mathpzc{w}_0^{(2)}\right)t}\right] - \varphi_-^{(2)}\varphi_+^{(2)}\left[1-e^{i\left(\mathpzc{w}_-^{(2)}-\mathpzc{w}_+^{(2)}\right)t}\right]\right\},\\ \\
\displaystyle
\phi(\pmb{k}-\pmb{\varkappa},t)\simeq-ia_0 \theta_-^{(2)}e^{-i\left(\mathpzc{w}_0^{(2)}-\varkappa_0\right)t}\left[1-e^{i\left(\mathpzc{w}_0^{(2)}-\mathpzc{w}_-^{(2)}\right)t}\right],\quad
f_+(\pmb{k}-2\pmb{\varkappa},t)\simeq -a_0\varphi_-^{(2)} e^{-i\left(\mathpzc{w}_+^{(2)}-2\varkappa_0\right)t}\left[1-e^{i\left(\mathpzc{w}_+^{(2)}-\mathpzc{w}_-^{(2)}\right)t}\right].
\end{array}
\end{eqnarray}While the  angles
\begin{eqnarray}\label{mixingangleweakapp2}
\theta_-^{(2)}\simeq\left.\frac{\phi(k-\varkappa)}{if_-(k)}\right\vert_{\omega=\mathpzc{w}_-^{(2)}}=\frac{g m_*^2\sqrt{I}}{8\sqrt{2\pi}\varepsilon_{\pmb{k}-\pmb{\varkappa}}\varkappa_0\left(\varepsilon_{\pmb{k}-\pmb{\varkappa}}+\varkappa_0-\omega_{\pmb{k}}\right)},\  \theta_0^{(2)}\simeq\left.\frac{f_-(k)}{i\phi(k-\varkappa)}\right\vert_{\omega=\mathpzc{w}_0^{(2)}}=\frac{g m_*^2\pi\sqrt{ I}}{2\sqrt{2\pi}\omega_{\pmb{k}}\varkappa_0\left(\varepsilon_{\pmb{k}-\pmb{\varkappa}}-\varkappa_0-\omega_{\pmb{k}}\right)},
\end{eqnarray}describe the respective ALP-photon mixing, the remaining  ones  are associated with  the  oscillations   between  photons with  different helicities. 
These can be approached by
\begin{eqnarray}
\varphi_-^{(2)}\simeq-\left.\frac{f_+(k-\varkappa)}{if_-(k)}\right\vert_{\omega=\mathpzc{w}_-^{(2)}}=\frac{\pi_0}{\omega_{\pmb{k}-2\pmb{\varkappa}}\left(\omega_{\pmb{k}}-\omega_{\pmb{k}-2\pmb{\varkappa}}-2\varkappa_0\right)},\quad \varphi_+^{(2)}\simeq\left.\frac{f_-(k)}{if_+(k-2\varkappa)}\right\vert_{\omega=\mathpzc{w}_+^{(2)}}=\frac{\pi_0}{\omega_{\pmb{k}}\left(\omega_{\pmb{k}}-\omega_{\pmb{k}-2\pmb{\varkappa}}-2\varkappa_0\right)}.
\end{eqnarray} Observe that when the approximation  $(\omega_{\pmb{k}}-\varkappa_0)^2\gg m^2+m_*^2$  is taking into account,  the expression of  
$\varepsilon_{\pmb{k}-\pmb{\varkappa}}$ [Eq.~(\ref{treelevelsoldutionmassivemodes})]  involved  in (\ref{mixingangleweakapp2}) approaches 
to $\varepsilon_{\pmb{k}-\pmb{\varkappa}}\approx\omega_{\pmb{k}}-\varkappa_0+(m^2+m_*^2)/[2(\omega_{\pmb{k}}-\varkappa_0)]$.

In the following, we confine ourselves to the flavor-like electromagnetic waves that are elastically scattered. To this end we re-express the dispersion relations 
[Eqs.~(\ref{photonlikesolution}) and (\ref{cirmodes1})]  in terms of the mixing angles:
\begin{eqnarray}
\mathpzc{w}_+^{(1)}(\pmb{k})=\omega_{\pmb{k}}-\frac{\pi_3}{2\omega_{\pmb{k}}}-\frac{1}{2}\frac{m^2-m_*^2}{\omega_{\pmb{k}}+\varkappa_0}\theta_+^{(1)}\theta_0^{(1)},\quad \mathpzc{w}_-^{(2)}(\pmb{k})=\omega_{\pmb{k}}-\frac{\pi_3}{2\omega_{\pmb{k}}}-\frac{1}{2}\frac{m^2+m_*^2}{\omega_{\pmb{k}}-\varkappa_0}\theta_-^{(2)}\theta_0^{(2)}.
\end{eqnarray}  Since we  assume that $\omega_{\pmb{k}}\gg\frac{\pi_3}{2\omega_{\pmb{k}}}-\frac{1}{2}\frac{m^2\mp m_*^2}{\omega_{\pmb{k}}\pm\varkappa_0}\theta_\pm^{(1,2)}\theta_0^{(1,2)}$,  one can
write the relevant  flavor-like electromagnetic components in the following form
\begin{eqnarray}\label{mejorsdr}
f_+(\pmb{k},t)&\simeq& a_0 e^{-i\omega_{\pmb{k}} t}\left\{1-2\theta_+^{(1)}\theta_0^{(1)} \sin^2\left(\frac{1}{4}\frac{m^2-m_*^2}{\omega_{\pmb{k}}+\varkappa_0}t\right)-2\varphi_+^{(1)}\varphi_-^{(1)}\sin^2\left(\frac{1}{2}\Delta \mathrm{M}_+ t\right)\right.\nonumber\\
&&+\left.i\left[\frac{\pi_3}{2\omega_{\pmb{k}}}t-\varphi_+^{(1)}\varphi_-^{(1)}\sin\left(\Delta\mathrm{M}_+ t\right)+\frac{\theta_+^{(1)}\theta_0^{(1)}}{2}\frac{m^2-m_*^2}{\omega_{\pmb{k}}+\varkappa_0}t-\theta_+^{(1)}\theta_0^{(1)}\sin\left(\frac{1}{2}\frac{m^2-m_*^2}{\omega_{\pmb{k}}+\varkappa_0}t\right)\right]\right\},\\
\label{cmejorsdrd}
f_-(\pmb{k},t)&\simeq& a_0 e^{-i\omega_{\pmb{k}} t}\left\{1-2\theta_-^{(2)}\theta_0^{(2)} \sin^2\left(\frac{1}{4}\frac{m^2+m_*^2}{\omega_{\pmb{k}}-\varkappa_0}t\right)-2\varphi_+^{(2)}\varphi_-^{(2)}\sin^2\left(\frac{1}{2}\Delta\mathrm{M}_-\ t\right)\right.\nonumber\\&&
+\left.i\left[\frac{\pi_3}{2\omega_{\pmb{k}}}t-\varphi_+^{(2)}\varphi_-^{(2)}\sin\left(\Delta \mathrm{M}_-\  t\right)+\frac{\theta_-^{(2)}\theta_0^{(2)}}{2}\frac{m^2+m_*^2}{\omega_{\pmb{k}}-\varkappa_0}t+\theta_-^{(2)}\theta_0^{(2)}\sin\left(\frac{1}{2}\frac{m^2+m_*^2}{\omega_{\pmb{k}}-\varkappa_0}t\right)\right]\right\},
\end{eqnarray}  where  only the leading terms  have been withheld. Note that  the following abbreviation $\Delta \mathrm{M}_{\pm}\equiv\omega_{\pmb{k}}-\omega_{\pmb{k}\pm2\pmb{\varkappa}}\pm 2\varkappa_0$
 has been used.

 \subsection{Conversion probabilities\label{conversionprosect}}

The contributions  proportional to $\theta_+^{(1)}\theta_0^{(1)}$, $\theta_-^{(2)}\theta_0^{(2)}$ and   $\varphi_+^{(1,2)}\varphi_-^{(1,2)}$ in Eqs.~(\ref{mejorsdr})  and (\ref{cmejorsdrd})
are  perturbative corrections to the leading order term $\sim e^{-i\omega_{\pmb{k}} t}$.
In correspondence, one   can express the relevant parts of the photon wave functions of the problem as follows
\begin{equation}\label{especial2}
f_\pm(\pmb{k},t)=\sqrt{ \frac{4\pi}{2\mathpzc{w}_\pm}}\mathpzc{A}_\pm(\pmb{k},t)e^{-i\mathpzc{w}_\pm t},
\end{equation}where  the  normalization factor $a_0=\sqrt{4\pi/2\mathpzc{w}_\pm}$ has been chosen. The respective amplitudes  of the waves approach  to
\begin{eqnarray}
\mathpzc{A}_+(\pmb{k},t)\approx e^{-i\theta_+^{(1)}\theta_0^{(1)}\sin\left(\frac{1}{2}\frac{m^2-m_*^2}{\omega_{\pmb{k}}+\varkappa_0} t\right)-i\varphi_+^{(1)}\varphi_-^{(1)}\sin\left(\Delta\mathrm{M}_+t\right)-2\theta_+^{(1)}\theta_0^{(1)}\sin^2\left(\frac{1}{4}\frac{m^2-m_*^2}{\omega_{\pmb{k}}+\varkappa_0} t\right)-2\varphi_+^{(1)}\varphi_-^{(1)}\sin^2\left(\frac{1}{2}\Delta\mathrm{M}_+ t\right)},\\
\mathpzc{A}_-(\pmb{k},t)\approx e^{-i\theta_-^{(2)}\theta_0^{(2)}\sin\left(\frac{1}{2}\frac{m^2+m_*^2}{\omega_{\pmb{k}}-\varkappa_0} t\right)-i\varphi_+^{(2)}\varphi_-^{(2)}\sin\left(\Delta\mathrm{M}_-t\right)-2\theta_-^{(2)}\theta_0^{(2)}\sin^2\left(\frac{1}{4}\frac{m^2+m_*^2}{\omega_{\pmb{k}}-\varkappa_0} t\right)-2\varphi_+^{(2)}\varphi_-^{(2)}\sin^2\left(\frac{1}{2}\Delta\mathrm{M}_- t\right)}.\label{resonantamplitudee}
\end{eqnarray}The substitution of  Eqs.~(\ref{especial2})-(\ref{resonantamplitudee}) into Eq.~(\ref{chiralityexpansio}) allows us to  analyze
the part of the probe beam which is elastically scattered. The resulting electromagnetic wave  involves the  effects coming from an ALP, a fact to be   exploited
in the search of this  weakly interacting particle.

Clearly, the square of $\mathpzc{A}_\pm(\pmb{k},t)$ provides the survival probability for an incoming photon
with positive/negative  helicity $\mathpzc{P}_{\gamma_\pm\to\gamma_\pm}(\pmb{k},t)=\mathpzc{A}_\pm^*(\pmb{k},t)\mathpzc{A}_\pm(\pmb{k},t)$. The resulting  expressions
are intrinsically associated with the  exponentials responsible for the damping  of the corresponding electromagnetic waves due to both  the photo-production
of an ALP and the  generation  of Raman-like  photons. Since the respective  exponents are extremely  small, the terms proportional to $\theta_\pm\theta_0$  define
the  photo-production probabilities of an ALP in the field of a strong wave. Explicitly
\begin{equation}\label{exffdssASFG}
\mathpzc{P}_{\gamma_\pm \to \phi_\pm }\simeq \frac{g^2 I m_*^4(\omega_{\pmb{k}}\pm\varkappa_0)}{2\omega_{\pmb{k}} \varkappa_0^2\left(m^2\mp m_*^2\right)^2}\sin^2\left(\frac{1}{4}\frac{m^2\mp m_*^2}{\omega_{\pmb{k}}\pm\varkappa_0}t\right),
\end{equation} where the following abbreviations $\gamma_\pm\equiv f_\pm(k)$, $\phi_\pm\equiv\phi(k\pm\varkappa)$ have been introduced.  Note that
both probabilities $\mathpzc{P}_{\gamma_\pm \to \phi_\pm}$  vanish   identically when $g\to0$. It is worth mentioning  that the following limit $\lim_{t\to\infty}\mathpzc{P}_{\gamma_\pm \to \phi_\pm}(t)/t=\mathpzc{R}_{\pm}$
provides the conversion rates in a pure  monochromatic plane wave. Considering  the relation $\pi\delta(x)=\lim_{\tau\to\infty}\sin^2(x\tau)/(x^2\tau)$
we find that
\begin{equation}
\mathpzc{R}_{\pm}=\frac{g^2m_*^4I\pi}{8\omega_{\pmb{k}}\varkappa_0^2}\delta\left(m^2\mp m_*^2\right).\label{rate1}
\end{equation}Manifestly, Eq.~(\ref{rate1})  shows that only the resonant process  can  occur in a  monochromatic plane wave, a fact  which  verifies the  statement  written  above
Eq.~(\ref{negmassmod}).

The  rate $\mathpzc{R}_{\pm}$ coincides with the one obtained from the standard  perturbation theory when the involved fields are canonically light-front-quantized \cite{selym}. 
Its  singularity at $m=m_*$ is an outcome of considering an infinity  interacting time. This fact  motivates us to investigate the realistic case where the field of the wave 
[Eq.~(\ref{externalF})] has a finite pulse length. In such a case it is expected that the Dirac delta in Eq.~(\ref{rate1}) be smeared out to a smooth function. The formalism 
developed in this section provides   evidences that this certainly takes place.

Now, the persistence probabilities also contain terms  proportional to $\sim\varphi_+^{(1,2)}\varphi_-^{(1,2)}$  which take into account the   generation of Raman-like
waves. Such terms reproduce the general expression for the probability  found in  \cite{Villalba-Chavez:2013txu}.  We combine  the respective outcomes
to    express   the total photo-production  probability of Raman-like waves as
\begin{eqnarray}\label{ponegtrans}
\begin{array}{c}\displaystyle
\mathpzc{P}_{\gamma\to\gamma^\prime}=\mathpzc{P}_{\omega\to\omega+2\varkappa_0}+\mathpzc{P}_{\omega\to\omega-2\varkappa_0},\\ \\
\displaystyle
\mathpzc{P}_{\omega\to\omega\pm2\varkappa_0}=\frac{4\pi_0^2}{\omega_{\pmb{k}}\omega_{\pmb{k}\pm2\pmb{\varkappa}}}\frac{\sin^2\left(\frac{1}{2}\left[(\omega_{\pmb{k}}-\omega_{\pmb{k}\pm2\pmb{\varkappa}}\pm 2\varkappa_0\right] t\right)}{\left(\omega_{\pmb{k}}-\omega_{\pmb{k}\pm2\pmb{\varkappa}}\pm 2\varkappa_0\right)^2},
\end{array}
\end{eqnarray}where the expressions for $\pi_0$ can be found in Eq.~(\ref{circualrformfactor}).
We remark that the expression above applies whenever the condition  $k\varkappa\simeq0$ is fulfilled  [see comment  below Eq.~(\ref{linerraman})].
So, it can be used in  the case in which  both lasers propagate quasi-parallelly, i.e.,   when  $k\varkappa\approx\omega_{\pmb{k}}\varkappa_0\theta^2/2\ll1$ with $\theta$
denoting the collision angle [$\theta\ll1$]. As a consequence,  the  conversion probability, resulting from the substitution of Eq.~(\ref{circualrformfactor}) into Eq.~(\ref{ponegtrans}), is given by
\begin{eqnarray}\label{strongfieldramman}
\mathpzc{P}_{\omega\to\omega\pm2\varkappa_0}\approx\frac{\alpha^2m_*^4\xi^{4}}{120^2 \pi^2m_0^4} \left\vert 1\pm2\frac{\varkappa_0}{\omega_{\pmb{k}}}\right\vert\sin^2\left(\frac{m_*^2}{\omega_{\pmb{k}}\pm2\varkappa_0}t\right).
\end{eqnarray}It is   opportune to emphasize that  Eq.~(\ref{strongfieldramman}) applies for   $\omega_{\pmb{k}}>2\varkappa_0$ or $2\varkappa_0>\omega_{\pmb{k}}$. 
In this context the resonant mass approaches to $m_*\simeq \theta(\omega_{\pmb{k}}\varkappa_0)^{\nicefrac{1}{2}}$. 
Once Eq.~(\ref{strongfieldramman}) is established, one can estimate the number of Raman-like photons generated during the interaction by considering the relation $\mathpzc{N}=\mathpzc{N}_0\mathpzc{P}_{\gamma\to\gamma^\prime}$ where  $\mathpzc{N}_{0}$ denotes the number of incoming probe
photons per shot. A  positive detection of such inelastic waves  would constitute a strong signature  of the  nonlinearity of
the  quantum vacuum. Unfortunately, the probability associated with this process  is extremely small $\sim\theta^4$, and  even for the forthcoming
high-intensity laser  facilities, the generation of a single Raman-like photon seems to be  extremely   difficult to achieve. It is  convenient to remark that the production rate of
Raman-like waves vanishes identically when both  laser waves counterpropagate and the strong one  approaches to the strict   monochromatic
situation \cite{Villalba-Chavez:2013txu}.

\section{Exclusion limits \label{section4}}

Hereafter, we  ignore the optical effects resulting  from the  Raman-like waves,  and  focus on those  associated with  the  axion-photon conversion. In the field of a circularly polarized plane wave, the vacuum  behaves as a chiral medium rather than a biaxial crystal  \cite{Villalba-Chavez:2013txu}.
As a consequence, the rotation of the polarization plane and the ellipticity of  the outgoing probe beam [Eq.~(\ref{chiralityexpansio}) with Eqs.~(\ref{especial2})-(\ref{resonantamplitudee}) included]
are determined by the relative phase between the  propagating modes and the difference between the photon absorption coefficients, respectively. Consequently,
the ellipticity of our problem approaches \cite{selym}
\begin{equation}\label{ellipticity}
\psi(t) \approx\frac{1}{4}\left\vert\mathpzc{P}_{\gamma_- \to \phi_-}-\mathpzc{P}_{\gamma_+ \to \phi_+}\right\vert
\end{equation}with $\mathpzc{P}_{\gamma_\pm\to\phi_\pm}$ as given in Eq.~(\ref{exffdssASFG}).   However, when  evaluating  $\psi(t)$,  we have to  keep  in mind  that the  experiment
must  include an external field  which  approaches   our monochromatic model [Eq.~(\ref{externalF})].  In practice,  the monochromaticity   of the high-intensity laser wave can be implemented
by  choosing   an appropriate  experimental setup in which the laser-source emits a pulse  with an  oscillation period  $\sim\varkappa_0^{-1}$  much smaller than its
temporal length $\tau$, i.e.  $\varkappa_0\tau\gg1$. For  $t=\tau$,  it is   expected  that the main contribution to the ellipticity comes from the resonant
term as it is dictated by Eq.~(\ref{rate1}):
\begin{equation}
\psi(\tau)\approx\frac{1}{4}\mathpzc{P}_{\gamma_+\to\phi_+} = \frac{g^2 I m_*^4(\omega_{\pmb{k}}+\varkappa_0)}{8\omega_{\pmb{k}} \varkappa_0^2\left(m^2-m_{*}^2\right)^2}\sin^2\left(\frac{1}{4}\frac{m^2- m_{*}^2}{\omega_{\pmb{k}}+\varkappa_0}\tau\right)\label{resonatellipticity},
\end{equation}where $m_*$ is  the resonant mass [Eq.~(\ref{resonancemass})]. At this point it is worth mentioning  that Eq.~(\ref{resonatellipticity})  applies whenever the condition
$(\omega_{\pmb{k}}+\varkappa_0)^2\gg2\epsilon m_*$ is fulfilled. If $\epsilon\ll 1\ \rm eV$ and $m_*\sim 1\ \rm eV$, we can then  restrict ourselves to the  case in which $\omega_{\pmb{k}}>\varkappa_0$
with $\omega_{\pmb{k}},\varkappa_0\sim 1\ \rm eV$, i.e. optical laser waves. Note that $\psi(\tau)$ is maximized when  the trigonometric argument is very small,  in which  case we find that
\begin{equation}
\psi(\tau)\approx\frac{1}{128}g^2\frac{I_c}{m_0^2}\frac{m_*^4 }{\omega_{\pmb{k}}(\omega_{\pmb{k}}+\varkappa_0)}\xi^2\tau^2. \label{resonantellypticity}
\end{equation}In this  expression  $I_c=m_0^4/(4\pi e^2)\approx4.6\times 10^{29}\ \rm W/cm^2$ denotes  the critical intensity, with $m_0$ and $\vert e\vert$ the  electron
mass and absolute charge, respectively. The square of the  intensity parameter $\xi^2=m_0^2 I/(\varkappa_0^2 I_c)$ is as  defined in Eq.~(\ref{direcpimunu}). So, near  resonance, an enhancement of the
ellipticity could occur as the product $\xi \tau$ increases.

The situation is   different for the angle $\vartheta(\tau)$  by which the polarization plane  is  rotated.  Whenever the high-intensity laser wave  approaches to  our monochromatic model,  we  find
\begin{equation}\label{rotationangle}
\vartheta(\tau)\approx\frac{1}{2}(\mathpzc{v}_--\mathpzc{v}_+)\omega_{\pmb{k}}\tau-\frac{g^2m_*^4 I(\omega_{\pmb{k}}+\varkappa_0)}{16\omega_{\pmb{k}}\varkappa_0^2\left(m^2-m_{*}^2\right)}\sin\left(\frac{1}{2}\frac{m^2-m_{*}^2}{\omega_{\pmb{k}}+\varkappa_0} \tau\right).
\end{equation}Here    $\mathpzc{v}_{+}$ and $\mathpzc{v}_{-}$ are  the phase velocities of the corresponding propagating  modes [Eq.~(\ref{phasevelocitycircular})].
The  resulting expression also applies whenever  the condition  $\omega_{\pmb{k}}>\varkappa_0$ is satisfied.
Note that the first term in Eq.~(\ref{rotationangle})  becomes   dominant  when  the resonance  is not reached.
On the contrary, when the argument of the trigonometric function  is very small,   the resonance contribution  in Eq.~(\ref{rotationangle})  vanishes identically and  the rotated angle
is simply    determined by
\begin{equation}\label{resonantangle}
\vartheta(\tau)\approx\frac{1}{2}\mathpzc{v}_-\omega_{\pmb{k}}\tau=\frac{1}{64}g^2\frac{I_c}{m_0^2}\frac{m_*^2}{\omega_{\pmb{k}}}\xi^2\tau.
\end{equation}Note that, Eq.~(\ref{resonantellypticity}) exceeds Eq.~(\ref{resonantangle})   by a factor
$\sim m_*^2 \tau/(\omega_{\pmb{k}}+\varkappa_0)$  as $\tau\to\infty$. Therefore, near resonance, the detection of  the ellipticity  seems to be more feasible  than the rotation
of the polarization plane.  This is the main   difference  between our laser-based setup and  investigations based on dipole magnets, where  the  opposite is true.

Now,  we wish to particularize Eqs.~(\ref{resonatellipticity})-(\ref{resonantangle}) to the case in which the collision is head-on, i.e, $\pmb{k}\cdot\pmb{\varkappa}=-\omega_{\pmb{k}}\varkappa_0$.  
Formally, the monochromaticity of our high-intensity laser wave [Eq.~(\ref{externalF})] implies to work in the limit of an  infinite pulse length \cite{jackson}. However, in practice,  
this is  a finite quantity and the monochromaticity  is guaranteed--up to certain limit--when the  strong   wave  is  characterized  by a relatively long pulse, i.e., $\tau\gg T$ with  
$T=2\pi\varkappa_0^{-1}$  the oscillating period.  The previous condition is satisfied 
by  choosing  the envisaged parameters  associated with OMEGA EP laser system \cite{OMEGA}  at Rochester, USA.  This system  will consist of four beamlines, two of which  capable of operating with  
a pulse-width range of $1-100\ \rm ps$ at central wavelength $\lambda_0\simeq1053\ \rm  nm$,
i.e., $\varkappa_0\simeq1.17\ \rm eV$.   For a pulse width  $\tau\simeq 1\ \rm ps$, the system will produce a power of the order of  $\sim 1\ \rm  PW$, i.e., $1\ \rm k J$ of pulse energy in
$1\ \rm ps$. Note  that in this setup the  product  $\varkappa_0\tau\sim 10^3 \gg1$,  which justifies   its
use  in  our monochromatic approach. We should also mention that the focal spot of the short-pulse beams is $80\%$  of the energy in a   $\sim 10\ \mu\mathrm{m}$-radius  spot, producing ultrahigh 
intensities $I$ exceeding the value  $2\times10^{20}\ \rm  W/cm^2$ corresponding to  $\xi\gtrsim10$.  We  can  suppose, in addition,  that the  experiment is  carried  out by coupling out a fraction 
of the strong wave  whose frequency is doubled [$\omega_{\pmb{k}}=2\varkappa_0$]  and  is  used as the probe beam. This guarantees the  necessary synchronization in the collision and allows us to 
study a resonant mass  $m_*\approx 3.3\ \rm eV$.  The exclusion limits are then determined   by requiring  that no significant signals are  detected at  certain  confidence level  neither in the 
ellipticity [Eq.~(\ref{resonatellipticity})] nor in the rotation angle [Eq.~(\ref{rotationangle})]. Searches of ALPs in a strong background laser field have not been carried out yet. However, in 
the optical regime of other laser-based experiments, sensitivities  of the order  of $\sim 10^{-10}\ \rm rad$  have been established \cite{Muroo_2003}.  Taking this value as reference,  a negative 
result in the  search  of the ellipticity  [Eq.~(\ref{resonantellypticity})] would  constrain $g\lesssim  1.3  \times 10^{-6}\ \mathrm{GeV}^{-1}$ near resonance. The resulting upper bound improves  
by  two orders of magnitude the constrains reported  in  \cite{Dobrich:2010hi,Dobrich:2010ie} by  using  the technical specification of the  POLARIS system \cite{POLARIS}. However, it roughly  remains 
two orders of magnitude  greater than the best laboratory constraint \cite{Ehret:2010mh,Ehret:2009sq}.

A   more stringent upper bound could by achieved by taking into account  the envisaged experimental parameters of  ELI and  XCELS projects. These ultra-high-intensity laser 
systems are planned to deliver a power of  $\sim 1\ \rm EW$, with  $\xi\approx 1.54\times 10^3$ [$I\approx10^{25}\ \rm W/cm^2$] and central  frequency $\varkappa_0\simeq1.55\ \rm eV$. For a temporal 
extension of $\tau\simeq 15 \ \rm fs$ this would not satisfy the  monochromaticity condition as  well as  the OMEGA EP facility. However, a  first   estimate   may  be carried out.  In  fact, by  
choosing the optical probe wave  as a fraction of the main laser  beam with $\omega_{\pmb{k}}=2\varkappa_0$ and by keeping the geometry of the collision, it is found that the   upper bound 
$g\lesssim 3.8\times 10^{-7}\ \rm GeV^{-1}$  applies for  a resonant mass $m_*\approx 4.4\ \rm eV$. We  emphasize that the outcome of this  analysis is  of particular importance  as it  may allows us  
to establish   the  extent  to which the monochromatic model correctly  describes  the phenomenology in these  ultra-short laser pulses, through   comparisons  with  more realistic models. In this 
context,  it is  worth mentioning that  the  order of magnitude of  our  exclusion limit  coincides with the one given  in \cite{selym}\  [$g\lesssim 1.8\times 10^{-7}\ \rm GeV^{-1}$],   established    
by considering  the external laser field  as  a  Gaussian pulse.

\begin{wrapfigure}{l}{0.45\textwidth}
\includegraphics[width=.45\textwidth]{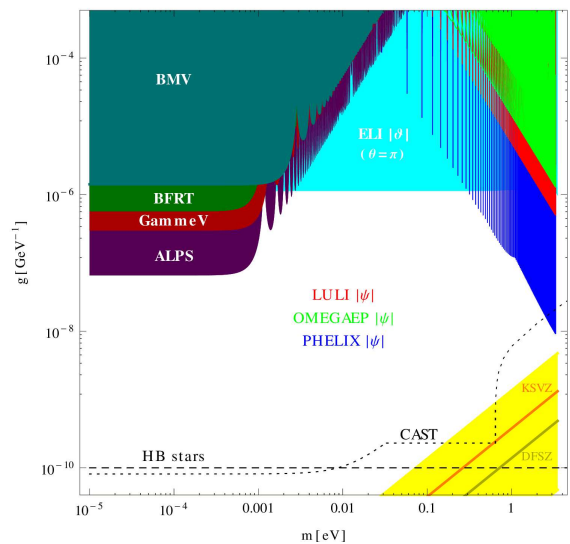}
\caption{\label{fig:mb001}(color online). Constraints for pseudoscalar  ALPs  of  mass $m$ and coupling constant $g$ obtained from a plausible polarimetric setup assisted by an intense circularly polarized 
laser field.   Multiple resonant peaks are displayed.  They were obtained by varying the collision angle [$\theta=1^\circ,2^\circ,3^\circ,\ldots,180^\circ$] and  by considering $\omega_{\pmb{k}}=2\varkappa_0$. 
Also shown  are the predictions of the axion models  with $\vert E/N-1.95\vert=0.07-7$ (the notation of this formula is in accordance with Ref. \cite{CAST}). 
The constraint resulting from the Horizontal Branch (HB) stars (dashed line) are  displayed as well. Further exclusion regions (shaded areas in the upper left corner) provided by  different experimental 
collaborations dealing with the Light Shining Through a Wall mechanism have also  been included. The  limit resulting from the solar monitoring of a plausible ALP flux \cite{CAST} is indicated by a dotted line. 
We remark  that the upper bound resulting from such an experiment strongly oscillates in the mass region  $0.4 \ \mathrm{eV}\leqslant m\leqslant0.6\ \rm eV$. This oscillating pattern 
has been replaced by  the exclusion limit $g\leqslant2.3\times 10^{-10}\ \rm GeV^{-1}$, established in \cite{CAST}  at $2\sigma $ confidence level. }
\end{wrapfigure}

Eq.~(\ref{resonantellypticity})  shows  that a laser pulse  with  a moderate intensity but with a large pulse length  can also be a sensitive  probe for pseudoscalar ALPs.  We investigate this situation   by  
choosing  a set of  parameters  associated with the Petawatt High-Energy Laser for heavy Ion eXperiments (PHELIX) \cite{PHELIX},  currently under operation in 
Darmstadt, Germany. In the nanosecond frontend, PHELIX operates with an infrared wavelength $\lambda_0\simeq1053\ \rm nm$ [$\varkappa_0\simeq1.17\ \rm eV$] and can reach a maximum intensity 
$I\simeq 10^{16}\  \rm W/cm^2$, corresponding to $\xi\simeq 6.4\times 10^{-2}$ in a pulse length $\tau\simeq20 \ \rm ns$. This large value of $\tau$  compensates the relative smallness of $\xi$, 
making  the product $\xi \tau \sim 10^{4}\ \rm eV^{-1}$  three orders of magnitude greater than the value resulting from ELI. As for the previous cases, we  suppose that the probe beam  
is an optical laser obtained  by coupling out a fraction of the strong laser whose  frequency is  shifted  to $\omega_{\pmb{k}}=2\varkappa_0=2.34\ \rm eV$ afterwards. By taking a sensitivity 
level of the order of $\sim 10^{-10}\ \rm rad$ we find that the upper limit $g\lesssim 9.1\times 10^{-9}\ \rm GeV^{-1}$ applies at  $m_*\simeq 3.3\ \rm eV$. 

Our exclusion regions  are given  in Fig.~\ref{fig:mb001}. The outcomes  in the upper right corner (blue, green and red) were derived  by considering an optical  experiment designed to 
detect a change in the ellipticity.  Clearly, the figure  shows how the parameter space to be excluded in the ($g,m$)-plane  increases  as different   collision angles are chosen 
[$\theta=1^\circ,2^\circ,3^\circ,\ldots,180^\circ$]. According to Eq.~(\ref{resonancemass}) and (\ref{resonatellipticity}),  each angle determines a resonant mass at which the  signal is maximized. 
A   set of  different resonant peaks translates into an exclusion comb which  depends  on the strong field source. 
The upper limit for   the  specification of the long high-energy pulse of  $400\ \rm J$   at the Laboratoire pour l'Utilisation des Lasers Intenses (LULI) \cite{LULI}--currently in operation at Palaiseau, France--can be seen 
as well. Similarly to OMEGAEP and PHELIX, the  nanosecond facility at  LULI(2000) system operates with a central frequency $\varkappa_0\simeq1.17 \ \rm eV$, but its  pulse length can reach   
$\tau\simeq 1.5 \ \rm ns$ for an  intensity of  $I\simeq6\times 10^{14}\ \rm W/cm^2$ [$\xi\simeq2\times 10^{-2}$]. 
For comparison, the prediction resulting from the hadronic models of Dine-Fischler-Srednicki-Zhitnitskii 
(DFSZ) \cite{Dine:1981rt,zhitnitskii}  and Kim-Shifman-Vainshtein-Zakharov  (KSVZ)  \cite{kim,shifman} axions have been included. Furthermore, the upper bound from the search for solar  axions 
\cite{CAST}  is indicated by a dashed line.  

Summing up, Fig.~\ref{fig:mb001} shows  that  high-precision polarimetric experiments assisted by the field of a high-intensity laser wave  could provide a sensitive probe 
for pseudoscalar ALPs in region of  masses for  which a laboratory setup   based on  dipole magnets provides less stringent limits. As is clear from the plot,  our upper bounds are  excluded 
by the constraint  resulting from considerations of stellar energy  loss due to the axion production in the  horizontal branch (HB) stars \cite{Raffelt:2006cw}.  However,  this kind of  
constraint must be considered with certain care  because there are macroscopic quantities such as    temperature and  density of the start,  whose inclusions can attenuate the limit  
 significantly  \cite{Gies:2007ua,evading}.  This renders well-controlled laboratory searches of ALPs --as the present proposal and the ones dealing with Light Shining Through 
a Wall setups--crucially important to complement   astro-cosmological studies.

\section{Summary and outlook}

In this article, the  mixing of  photon with an ALP  mediated by  a  strong  circularly polarized monochromatic
plane wave has been analyzed.   The effects   resulting from   the  interaction between a small-amplitude electromagnetic wave
and the vacuum polarized by the field of a strong  wave  were also considered.  In correspondence, the low energy behavior
of the polarization tensor in the field of a plane wave of arbitrary shape was  determined. We have seen that
the specific shape  of  the external wave  makes  the  conversion processes conceptually more  complex  than in the case where the mixing
is assisted by  dipole magnets. However, the inherent simplifications   of  the monochromatic paradigm  compared to  waves  modulated by
particular  profiles,  allows   some particular  aspects  of the  ALP-photon oscillations to be  establish  in a concise way.

A detailed perturbative treatment has been  implemented  for determining  the flavor-like fields as well as  the relevant dispersion  relations.
It was found that, in a circularly polarized monochromatic plane wave and at energies below the scale specified by the electron mass,   the  pure QED vacuum  behaves as a nonbirefringent medium.
The incorporation of  ALP-photon coupling induces  a tiny birefringence and dichroism in the vacuum.  The corresponding
expressions for the ellipticity and the angular  rotation  of the polarization plane were used to impose exclusion limits on the ALPs attributes.
We have  also shown  that the most stringent constraints  on  the coupling constant  are  in the vicinity
of resonant  masses which depend  on  the  frequency  of  both  laser fields.

While  our research  does not cover  all plausible experimental setups, the general expressions  obtained in Sec. \ref{svpfgpw} certainly  apply
to other external  configurations of laser fields as well.  As a consequence,  they can be used  in cases where the strong  plane wave  is, for instance,
a bichromatic wave or a Gaussian pulse. Both  problems  are  expected to be  more cumbersome and  procedures other than  the one used in
this work, may be required. Besides,  the  analysis in such field configurations might reveal whether the generation of Raman-like  waves is  favored
when  both lasers counterpropagate. If  so,  we will have at our disposal another mechanism   for  probing the nonlinear behavior of the quantum vacuum.
We plan to present  detailed studies  of these problems   in  forthcoming publications.

\vspace{0.005 in}
\begin{flushleft}
\textbf{Acknowledgments}
\end{flushleft}
\vspace{0.005 in}

I am very grateful to A. Di Piazza,  C. M\"{u}ller and B. D\"{o}brich,  for helpful discussions.  I would  also like to extend my
gratitude to John Farmer and C. M\"{u}ller for their  critical reading  and valuable comments on the manuscript.

\end{document}